\documentclass[11pt,reqno]{amsart}
\usepackage{hyperref}
\usepackage{url}
\usepackage{amssymb}
\usepackage[T1]{fontenc}
\usepackage{extarrows}
\usepackage{lmodern}
\usepackage[utf8]{inputenc}
\usepackage{ifthen}
\usepackage{mathtools}
\usepackage{xcolor}
\usepackage{amsmath}
\usepackage{booktabs}
\usepackage{array}
\usepackage{geometry}
\newtheorem{thm}{Theorem}
\newtheorem{coro}{Corollary}

\newtheorem{lem}{Lemma}
\newtheorem{prop}{Proposition}
\newtheorem{ex}{Example}
\newtheorem{rmk}{Remark}
\theoremstyle{definition}
\newtheorem{definition}{Definition}
\theoremstyle{remark}

\numberwithin{equation}{section}

\newcommand{\F}{\mathbb{F}}
\newcommand{\C}{\mathbb{C}}

\newcommand{\Z}{\mathbb{Z}}

\usepackage{tikz,enumerate}
\begin{document}

\title{Codes in multivariate $P$-polynomial association schemes}
\author{Minjia Shi}
\address{Key Laboratory of Intelligent Computing Signal Processing, Ministry of Education, School of Mathematical Sciences, Anhui University, Hefei 230601, China}
\email{smjwcl.good@163.com}
\author{Jing Wang}
\address{School of Mathematical Sciences, Anhui University, Hefei 230601, China}
\email{wangjing031010@163.com}
\author{Patrick Sol\'e}
\address{I2M (Aix Marseille Univ, CNRS), Marseilles, France}
\email{sole@enst.fr}
\begin{abstract}
We generalize the fundamental results of Delsarte thesis (1973) on codes in $P$-polynomial association schemes in the framework introduced by Bannai et al. (2025). We assume the scheme is weakly metric in the sense of (Sol\'e, 1989).
We give a Lloyd theorem on perfect codes in $P$-polynomial association schemes. We determine the rank of the outer distribution matrix and derive a linear relation on its columns. Similar results are derived for the outer distance matrix. Upper bounds on the covering radius of codes with known dual relation (resp. distance) spectrum size follow. A Lloyd-type relation is derived for the case of equality in the clique/coclique bound. This is relevant to diameter perfect codes. The width of the inner distribution and the external distance of the code are connected by a linear inequality. Equality in that bound yields completely regular codes.
\end{abstract}

\maketitle
\ \

\section{Introduction}
Delsarte's landmark 1973 PhD thesis laid the foundations for a major part of algebraic combinatorics \cite{D2}. As a researcher in an industrial lab, he was interested in the two main combinatorial structures relevant to Information Theory: codes and designs. To give bounds on their size he employed {\em linear programming}, a tool from combinatorial optimization \cite{Ch}. This allowed him to characterize the extremal objects meeting these bounds with equality (perfect codes, tight designs). The main innovation, at the time, was to derive these bounds in the context of {\it association schemes}, an algebraic structure coming from both  statistics \cite{Ba} and group theory \cite{BI,BCN}.
The applicability of these techniques, unfortunately, was limited to some particular association schemes
(metric schemes, Q-polynomial association schemes), which enjoy a very simple generator structure of their Bose-Mesner algebra. However, in many applications, association schemes that are not metric and/or not Q-polynomial are needed \cite{AGKP,NR,MW,NRT,qary,S3}. In \cite{S2}, was introduced the notion of {\em weakly metric} association scheme (WMAS), which was further developed in \cite{S1}. A WMAS is essentially the datum of an association scheme with a distance that is constant on the relations of the scheme. For instance every metric scheme is weakly metric for the geodetic distance on the assocaited distance regular graph. In 2024 the concept of m-distance regular graph, generalizing metric schemes and distance regular graphs, was introduced in Bernard et al. \cite{mreg}. In 2025, the notion of {\em multivariate} P-polynomial (and also Q-polynomial) association schemes was introduced by Bannai et al. \cite{B+}.

In a previous paper, we have extended the combinatorics of Q-polynomial schemes in \cite{D2} to multivariate Q-polynomial schemes \cite{SWS}. In the present paper, we extend the P-polynomial results of \cite{D2} to multivariate P-polynomial schemes, taking into account more recent developments like diameter perfect codes \cite{E} and the concept of width of the inner distribution \cite{width}. We begin by establishing a bridge between $m $-distance regular graphs and weakly metric association schemes. We show that for a suitable choice of monomial domain in a multivariate P-polynomial scheme there is a natural distance for which the scheme is weakly metric. Then we proceed to derive a Lloyd theorem in the multivariate setting. We define two outer distribution matrices depending on whether the columns are indexed by distances or by compositions. This is the main difference between metric and weakly metric schemes. We derive linear relations among the columns of these outer distribution matrices, using the external spectral support of the code. In some examples we are able to determine the matrix completely. We also derive a spectral upper bound on the covering radius in the spirit of the  classical Delsarte bound in the Hamming scheme \cite{D}. We obtain a Lloyd-type spectral characterization of equality in the clique-coclique bound. This result is new, even in the one variable case. We revisit the concept of width of a code and give an inequality connecting the width with the external distance. In case of equality the code is shown to be completely regular \cite{SS}.

The material is organized as follows. The next section  provides the preliminary notions and notation needed for the rest of the paper; polynomial weakly metric schemes, multivariate $P$-polynomial association schemes, $m$-distance-regular graphs. Section 3 collects examples of weakly metric association schemes. Section 4 derives a Lloyd theorem for perfect codes in P-polynomial association schemes. Section 5 studies the outer distribution matrix. The short section 6 focuses on equality in the clique-coclique bound. Section 7 studies width and completely regular codes. Section 8 concludes the article.
\section{Preliminaries}
\subsection{Weakly metric association schemes}
Let $X$ be a finite set with at least two elements and let $\mathcal R = \{R_0, R_1, \ldots, R_s\}$ be a family of $s + 1$ relations $R_i$ on $X$ for $s \geq 1$. Let $\mathcal{I}=\{0,1,\ldots,s\}$ denote the index set of these relations. The pair $\mathfrak{X}=(X,\mathcal R)$ is called a commutative association scheme with $s$ classes \cite{XSS} if the following conditions are satisfied:
\begin{enumerate}
\item[(A1)] The set $\mathcal R$ is a partition of $X^2$ and $R_0 = \{(x, x) \mid x \in X\}$ is the diagonal relation;
\item[(A2)] For $i = 0, 1, \ldots, s$, the inverse $R_i^{-1} = \{(y, x) \mid (x, y) \in R_i\} $ of the relation $R_i$ also belongs to $R$;
\item[(A3)] For any triple of integers $i, j, k = 0, 1, \ldots, s$, there exists a number $ p_{ij}^k = p_{ji}^k$ such that
$$ |\{z \in X \mid (x, z) \in R_i, (z, y) \in R_j\}| = p_{ij}^k $$
for all \( (x, y) \in R_k \).
\end{enumerate}
We say that an association scheme $\mathfrak{X}=(X,\mathcal R)$ is {\bf weakly metric} (shortly WMAS) for a distance $d$ if the distance is constant on the classes of the scheme.
( this is slightly more restrictive than the definition in \cite{S2} where $d$ is only assumed to be a quasi-distance). Formally, the scheme $\mathfrak{X}=(X,\mathcal R)$ with $s$ classes is weakly metric for $d$ if for all $(a,b)\in X^2$ and for all $i\in\{0,1,\ldots,s\},~aR_{i}b~\Rightarrow~d(a,b)=d(i)$.

For each relation $R_i$ ($i = 0, 1, \ldots, s$), we define its adjacency matrix $A_i$ by:
$$ A_i(x, y) =
\begin{cases}
1 & \text{if } (x, y) \in R_i, \\
0 & \text{if } (x, y) \notin R_i.
\end{cases} $$
The set of all complex linear combinations of these adjacency matrices forms a complex algebra:
\[
\mathcal{T}=\left\{\sum_{i=0}^s \alpha_i A_i \mid \alpha_i \in\mathbb{C}\right\},\]
known as the Bose-Mesner algebra of the association scheme.
The Bose-Mesner algebra admits two natural bases. The first is the set of adjacency matrices $\{A_i\}_{i=0}^s$, and the second is a unique basis of primitive idempotents $\{E_i\}_{i=0}^s$ satisfying:
\[
E_i^2 = E_i,~E_i E_j = 0\ (i \neq j),~\sum_{i=0}^s E_i = I_{|X|},
\]
where $I_{|X|}$ denotes the identity matrix of order $|X|$.
The linear transformations between these bases are given by \cite{eigenvalues}
\[
A_i=\sum_{j=0}^{s}P_i(j)E_j,~E_j=\frac{1}{|X|}\sum_{i=0}^{s}Q_j(i)A_i,~i=0,1,\ldots,s.
\]
The complex numbers \(P_i(0),P_i(1),\ldots,P_i(s)\) are the eigenvalues of \(A_i\), called the \textbf{first eigenvalues} of the association scheme, while \(Q_j(i)\) are called the \textbf{second eigenvalues}. The matrices \(P=(P_i(j))\) and \(Q=(Q_j(i))\) are called the \textbf{first eigenmatrix} and the \textbf{second eigenmatrix} respectively, and satisfy
\[
PQ=QP=|X|I_{s+1}.
\]
Let \(v_i=P_i(0),~ u_j=Q_j(0)\). It is easy to see that \(v_i\) is the valency of the relation \(R_i\), and \(u_j=\operatorname{rank}(E_j)\) is the multiplicity of the corresponding primitive idempotent.

The {\bf Lloyd polynomial} \cite{S1} in a weakly metric association scheme is defined by
\[
\Psi_{r}(j) = \sum_{d(k) \leq r} P_{k}(j).
\]

In a weakly metric scheme $\mathfrak{X}=(X, \mathcal R)$ with distance $d$, the {\bf dispersion function} $\Pi$ is defined by
\[
\Pi(j) = |\{i \in \{0, 1, \ldots, s\} \mid d(i) \leq j\}|.
\]

An {\bf e-perfect code} $C \subset X$ in $\mathfrak{X}=(X, \mathcal R)$ is then defined as a code satisfying the partition
\[
X=\coprod_{c \in C} B(c,e),
\]
where $\coprod$ means disjoint union, and $B(c,e)=\{y\in X \mid d(y,c)\le e\}$ is the {\bf ball of radius e} for the distance $d$ centered in $c.$
We are now ready to state the Lloyd Theorem in WMAS.

{\thm\label{Lloyd} $($\cite{S1}$)$ If there exists an $e$-perfect code in $(X, \mathcal R)$ then the Lloyd polynomial $\Psi_e(j)$  vanishes for at least $\Pi(e)-1$ indices $j.$ }

\subsection{Multivariate P-polynomial association schemes}
\begin{definition}
A \textbf{monomial order} $\leq$ on $\mathbb{C}[x_1, x_2, \ldots, x_m]$ is a relation on the set of monomials $x_1^{n_1} x_2^{n_2} \ldots x_m^{n_m}$ satisfying:
\begin{enumerate}
\item $\leq$ is a total order;
\item for monomials $u, v, w$, if $u \leq v$, then $wu \leq wv$;
\item $\leq$ is a well-ordering, i.e., any non-empty subset of the set of monomials has a minimum element under $\leq$.
\end{enumerate}
\end{definition}

Since each monomial $x_1^{n_1} x_2^{n_2} \ldots x_m^{n_m}$ can be associated to the tuple
$(n_1, n_2, \ldots, n_m) \in \mathbb{N}^m$, we use the same order and notation $\leq$ on
$$ \mathbb{N}^m = \{(n_1, n_2, \ldots, n_m) \mid n_i \text{ are non-negative integers}\}. $$
Note that for $m = 1$, there is a unique monomial order, which is the one associated to the natural ordering of the integers.

\begin{definition}
For a monomial order $\leq$ on $\mathbb{N}^m$, an $m$-variate polynomial $v(x_1, x_2, \ldots, x_m) = v(x)$ is said to be of \textbf{multi-degree} $n \in \mathbb{N}^m$ if $v(x)$ is of the form
$$ v(x) = \sum_{a \leq n} f_a x^a, \quad x^a = x_1^{a_1} x_2^{a_2} \ldots x_m^{a_m}, $$
with $f_n \neq 0$.
\end{definition}

Let $\epsilon_i \in \mathbb{N}^m$ be the tuple whose $i$-th entry is $1$ and all the other entries are $0$.

\begin{definition}
Let $\mathcal{D} \subset \mathbb{N}^m$ contain $\epsilon_1, \epsilon_2, \ldots, \epsilon_m$ and $\leq$ be a monomial order on $\mathbb{N}^m$. A commutative association scheme $\mathfrak{X}$ is called \textbf{m-variate P-polynomial} \cite{B+} on the domain $\mathcal{D}$ with respect to $\leq$ if the following three conditions are satisfied:

\begin{enumerate}
\item If $n = (n_1, n_2, \ldots, n_m) \in \mathcal{D}$ and $0 \leq n'_i \leq n_i$ for all $i$, then
$$ n' = (n'_1, n'_2, \ldots, n'_m) \in \mathcal{D}; $$
\item There exists a relabeling of the adjacency matrices of $\mathfrak{X}$:
\(\{A_i\}_{i\in \mathcal I}=\{A_n\mid n\in\mathcal{D}\}\) such that for all \(n\in\mathcal D\) we have
$$A_n=v_n(A_{\epsilon_1},A_{\epsilon_2},\ldots,A_{\epsilon_m}),$$
where $v_n(x)$ is an $m$-variate polynomial of multidegree $n$ with respect to $\leq$, and all monomials $x^c$ in $v_n(x)$ satisfy $c\in\mathcal D$;
\item For $i = 1, 2, \ldots, m$ and $a = (a_1, a_2, \ldots, a_m) \in \mathcal{D}$, the product $A_{\epsilon_i} A_{\epsilon_1}^{a_1} A_{\epsilon_2}^{a_2} \cdots A_{\epsilon_m}^{a_m}$ is a linear combination of
$$ \{A_{\epsilon_1}^{b_1} A_{\epsilon_2}^{b_2} \cdots A_{\epsilon_m}^{b_m} \mid b = (b_1, \ldots, b_m) \in \mathcal{D}, \, b \leq a + \epsilon_i\}. $$
\end{enumerate}
\end{definition}

{\rmk
For $m=1$, the above definition reduces to the classical notion of $P$-polynomial association scheme, since the monomial order on $\mathbb{N}$ is the usual order and \(A_n=v_n(A_1)\).}

We now introduce the notion of $m$-distance and $m$-distance regular graphs.
\subsection{m-distance regular graphs} \cite{mreg}
Let $G = (X, \Gamma)$ be a connected undirected graph with vertex set $X$ and edge set $ \Gamma$. We consider an $m$-partition $\{\Gamma_i \mid i = 1, 2, \ldots, m\}$ of $\Gamma$, i.e.
\[ \Gamma = \Gamma_1 \sqcup \Gamma_2 \sqcup \cdots \sqcup \Gamma_m,~ \Gamma_i \neq \emptyset, \]
where $\sqcup$ denotes disjoint union of sets. This can be interpreted as a colouring of the edges of $G$ with $m$ colours, with edges in $\Gamma_i$ coloured $i$.

In the following, a walk on $G$ is a finite ordered sequence of edges $\xi=(\gamma_1, \gamma_2, \ldots, \gamma_L)$ such that $\gamma_i$ and $\gamma_{i+1}$ share a common vertex. Denote $\gamma_i$ as $\gamma_i = (y_i, y_{i+1}) \in \Gamma$. The vertices $y_1$ and $y_{L+1}$ constitute the two endpoints of $\xi$. The endpoints of $\xi = \emptyset$ can be taken to be any $y \in X$. If for $i = 1, 2, \ldots, L$ the edge $\gamma_i$ has colour $c_i$, that is, $\gamma_i \in \Gamma_{c_i}$, then we say that the walk $\xi = (\gamma_1, \gamma_2, \ldots, \gamma_L)$ is of type $c = (c_1, c_2, \ldots, c_L)$.
\begin{definition}
Let $\xi = (\gamma_1, \gamma_2, \ldots, \gamma_L)$ be a walk on $G$. Its $m$-length $\ell_m(\xi)$ with respect to the $m$-partition $\{\Gamma_i \mid i = 1, 2, \ldots, m\}$ is the vector in $\mathbb{N}^m$ whose $i$-th coordinate is the number of edges in the walk of colour $i$, that is,
$$ \ell_m(\xi) = (|\{j \mid \gamma_j \in \Gamma_1\}|, |\{j \mid \gamma_j \in \Gamma_2\}|, \ldots, |\{j \mid \gamma_j \in \Gamma_m\}|). $$
\end{definition}

\begin{definition}
Let $\leq$ be a monomial order on $\mathbb{C}[x_1, \ldots, x_m]$. The \textbf{$m$-distance} $d_m$ between two vertices $x, y$ of a connected graph $G = (X, \Gamma = \Gamma_1 \cup \cdots \cup \Gamma_m)$ is defined as
$$ d_m(x, y) = \min_{\leq} \{ \ell_m(\xi) \mid \xi \text{ is a walk between } x, y \}. $$
\end{definition}

The $m$-distance is symmetric (i.e. $d_m(x, y) = d_m(y, x)$) because $G$ is undirected so that any walk from $x$ to $y$ gives rise to a walk from $y$ to $x$ when the order of edges is reversed. Note that $m$-distance is not a distance in the traditional sense as it is not a real-valued function (apart from the case $m = 1$, which corresponds to the usual graph distance). In the following, we denote by \(\Delta\) the set of all $m$-distances between pairs of vertices in $G$,
$$ \Delta = \{ \ell = (\ell_1, \ldots, \ell_m) \in \mathbb{N}^m \mid \exists x, y \in X \text{ s.t. } d_m(x, y) = \ell \}. $$
Note that $0 := (0, 0, \ldots, 0) \in \Delta$ since $d_m(x, x) = (0, 0, \ldots, 0)$ for all $x \in X$.\\

We define the distance matrices associated with the $m$-distance $d_m$ of a graph as follows.

\begin{definition}
Let $G = (X, \Gamma)$ be a graph with $m$-partition $\{\Gamma_i \mid i = 1, 2, \ldots, m\}$ and let \(\Delta\) be the set of $m$-distances of $G$ with respect to a monomial order $\leq$ on $\mathbb{N}^m$. For $\ell \in \Delta$, the \textbf{$\ell$-th $m$-distance matrix $A_\ell$} of the graph $G$ is the matrix whose columns and rows are labelled by elements of $X$ and whose entries satisfy
$$ (A_\ell)_{xy} =
\begin{cases}
1, & \text{if } d_m(x, y) = \ell, \\
0, & \text{otherwise}.
\end{cases}. $$
\end{definition}
The $m$-distance matrices $A_\ell$ are symmetric, since the $m$-distance is a symmetric function of the vertices. In the case $m = 1$, the function $d_1$ reduces to the length of a shortest path in $G = (X, \Gamma)$ and the matrices $A_\ell$ correspond to the usual $\ell$-th distance matrix of $G$.

\begin{definition}
A connected undirected graph is said to be \textbf{$m$-distance-regular} if there exists an $m$-partition $\{\Gamma_i \mid i = 1, 2, \ldots, m\}$ of its edges and a monomial order $\leq$ such that $\epsilon_1, \epsilon_2, \ldots, \epsilon_m$ are in the set \(\Delta\) of all $m$-distances in $G$ and the number $p^c_{ab}$ of vertices $z$ at $m$-distance $a = (a_1, \ldots, a_m)$ from a vertex $x$ and at $m$-distance $b = (b_1, \ldots, b_m)$ from a vertex $y$ given that $x$ and $y$ are at $m$-distance $c = (c_1, \ldots, c_m)$ does not depend on the choice of $x$ and $y$.
\end{definition}

\subsection{Polynomial schemes}
We shall say that a WMAS with $s$ classes is {\bf $r$-polynomial} if there exist $s+1$ points of $\C^r$  such that the first eigenvalues $P_k(i)$ are polynomials of degree $\le d(k)$ in $r$ variables evaluated in these points. Thus P-polynomial schemes in the sense of \cite{BI} are $1$-polynomial by \cite[Proposition 1.1. p. 189--190]{BI}. The {\bf index set} $S$ is the smallest finite set such that the above $s+1$ points are in $S^r.$ We will write, for any integer $N$ the interval of integers from zero to $N$ as $[[N]].$

{\bf Example:} Consider the Clark-Liang scheme $CL(15,2) $ \cite{S1}. This scheme is weakly metric for the modular distance. Concretely, denote by $C_i$ the 2-cyclotomic coset containing $i$ as its smallest element. This notation is familiar from the study of binary cyclic codes \cite{MS}. Write $X_1=C_1 \cup C_7,\,X_2=C_3,\,X_3=C_5.$
For $i=1,2,3$ declare that for $x,y \in X=\Z_{15}$ we have $x R_i y$ iff $x-y \in X_i.$  We have $d(1)=1,\, d(2)=d(3)=2.$ This scheme is not $P$-polynomial as the only adjacency matrix spanning the Bose-Mesner algebra is $A_1$ (Dimension computation in Magma \cite{Ma}). However, the Cayley graph on $\Z_{15}$ generated by $X_1$ has diameter $2<3.$  We conjecture that the scheme  $2$-variate $P$-polynomial in the sense of \cite{B+} with domain $\mathcal{D}=\{20,10,00,01\},$ and basis $(A_1,A_2).$
Indeed it can be checked that $4A_3=A_1^2-3A_1-6A_2-8I.$ Multiplying on the right by $J_i$ we get $4P_3(i)=P_1(i)^2-3P_1(i)-6P_2(i)-8.$ This entails that $CL(15,2) $ is $2$-polynomial for the points $(P_1(i),P_2(i))$ for $ i \in [[3]].$

A symmetric $m$-variate $P$-polynomial association scheme corresponds to an $m$-distance regular graph $G=(X,\Gamma)$. Its vector-valued distance function $d_m:X\times X\to\mathbb N^m$ satisfies the triangle inequality under a given monomial
order \cite[Theorem 3.10]{mreg}. Since the relation classes $R_\ell$ of the scheme are determined by the values of $d_m$, i.e., \(R_\ell=\{(x,y)\mid d_m(x,y)=\ell\}\), the function \(\widetilde d(x,y)=|d_m(x,y)|_1\) is constant on each relation class, where $|z|_1=|z_1|+\cdots+|z_m|$ for $z\in\mathbb R^m$.
We now give a sufficient condition under which the scheme constitutes a WMAS for $\widetilde d$.

\begin{definition}
A monomial order $\leq$ is said to be \textbf{$L^1$ compatible} with $\mathcal D$ if, for all $a\in\mathcal D$ and $b\in\mathcal D+\mathcal D$, we have \(a\leq b\Longrightarrow |a|_1\leq |b|_1\), where $\mathcal D+\mathcal D=\{u+v\mid u,v\in\mathcal D\}$, and the second $\leq$ denotes the usual order on the reals.
\end{definition}

{\prop \label{weakly metric} Keep the above notation, and assume that the monomial order $\leq$ is $L^1$ compatible with $\mathcal D$. The association scheme canonically associated with an $m$-distance-regular graph is weakly metric with respect to $\widetilde{d}$.}
\begin{proof}
By the above discussion it is clear that $\widetilde{d}$ is constant on the classes of the scheme. We only need to verify the triangle inequality. By \cite[Lemma 3.8]{mreg}, we have $d_m(x,y)\le d_m(x,z)+d_m(z,y)$ for all $x,y,z \in X$. The result follows then by $L^1$ compatibility.
\end{proof}

Define $d(\ell)=|\ell|_1$ for $\ell\in\mathcal D$, thus $\widetilde d(x,y)=d(\ell)$ whenever $(x,y)\in R_\ell$. For $0\leq j\leq s$, define the \textbf{spectral points}
\(p_j=(P_{\epsilon_1}(j),\ldots,P_{\epsilon_m}(j))\in\mathbb C^m\).

{\prop \label{L1-WMAS} Keep the above notation, and assume that the monomial order $\leq$ is $L^1$ compatible with $\mathcal{D}$, then the $m$-variate $P$-polynomial scheme $\mathfrak X$ is an $m$-polynomial WMAS with respect to $\widetilde d$.}
\begin{proof}
By Proposition \ref{weakly metric}, $\mathfrak X$ is weakly metric with respect to $\widetilde d$, where $d(n)=|n|_1$ for $n\in\mathcal D$.
Since $A_n=v_n(A_{\epsilon_1},\ldots,A_{\epsilon_m})$ and $A_{\epsilon_i}E_j=P_{\epsilon_i}(j)E_j$, we obtain \(P_n(j)=v_n(p_j),~n\in\mathcal D,~0\leq j\leq s\). These spectral points are distinct, because otherwise two columns of the first eigenmatrix would coincide.
	
Since $v_n$ has multidegree $n$, every monomial $x^b$ occurring in $v_n$ satisfies
$b\in\mathcal D$ and $b\leq n$, with a nonzero coefficient of $x^n$. As $0\in\mathcal D$, we have $n\in\mathcal D+\mathcal D$. Thus $L^1$ compatibility gives $|b|_1\leq |n|_1$, and hence \(\deg(v_n)=|n|_1=d(n)\). Therefore, the first eigenvalues admit $m$-variate polynomial representations of degree at most the corresponding distance,
which proves that $\mathfrak X$ is an $m$-polynomial WMAS.
\end{proof}

{\bf Examples:}
Many $L^1$ compatible monomial orders exist. For instance the {\em deg-lex} monomial order of \cite[eq.(41)]{mreg} satisfies this condition by definition. In particular Examples 4.2 and 4.3 of \cite{mreg} use that special monomial order. Example 4.1 allows for any monomial order \cite{B+,mreg}. It is a direct product of metric schemes. This situation covers the case of Lee metric over $\F_q$ (\S 3.3 direct product of ordinary $q$-gons) and of the scheme of \S 3.4 (direct product of bilinear forms graphs).

\subsection{Inner and outer distribution}
Let \(C\subseteq X\) be a nonempty subset. The \textbf{inner distribution} of \(C\) with respect to \(\{R_\alpha\}_{\alpha\in\mathcal I}\) is the vector
\(a(C)=(a_\alpha)_{\alpha\in\mathcal I}\), where  \(a_\alpha=\frac{1}{|C|}
|R_\alpha\cap(C\times C)|\). Equivalently, $a_\alpha$ is the average, over all $c\in C$, of the number of codewords in relation $R_\alpha$ with $c$.

The \textbf{outer distribution matrix} of \(C\) is the matrix \(B(C)\) whose rows are indexed by \(X\), whose columns are indexed by \(\mathcal I\), and whose \((x,\alpha)\)-entry is given by
\[
B(C)_{x,\alpha}=|R_\alpha\cap(\{x\}\times C)|=|\{c\in C:(x,c)\in R_\alpha\}|,~x\in X,\ \alpha\in\mathcal I.
\]
For a fixed \(x\in X\), let \(B(x)\) denote the row of \(B(C)\) corresponding to \(x\). Thus \(B(x)\) is the outer distribution of \(x\) with respect to \(C\).

We characterize \(C\subseteq X\) by its \textbf{characteristic vector} \(\chi_C\in \mathbb C^X\), defined by
\[
\chi_C(x) = \begin{cases} 1, & x\in C,\\ 0, & x\notin C. \end{cases}
\]
\(\chi_C\) is regarded as a column vector of length \(|X|\).

Recall the notion of an $M$-clique from Delsarte \cite[Sec.~3.3]{D2}. Let $M \subseteq \mathcal I$ with $0 \in M$, and write \(M^*=M \setminus \{0\},~ \overline{M}=\mathcal I \setminus M^*\). A nonempty subset $Y \subseteq X$ is called an \textbf{$M$-clique} if \(R_\alpha \cap (Y \times Y)=\emptyset,~\forall\, \alpha \in \mathcal I \setminus M\). Likewise, a nonempty subset $Z \subseteq X$ is called an \textbf{$\overline{M}$-clique} if \(R_\alpha \cap (Z \times Z) = \emptyset,~\forall\, \alpha \in M^*\).

\section{Some weakly metric schemes}
\subsection{Homogeneous metric}
Let $\mathbb{Z}_{2^k}$ be the ring of integers modulo $2^k$, where $k \geq 2$ is an integer.
For $n \geq 1$, $\mathbb{Z}_{2^k}^n$ denotes the set of $n$-tuples over $\mathbb{Z}_{2^k}$.
For an element $x \in \mathbb{Z}_{2^k}$, the \textbf{homogeneous weight} $W_{\mathrm{hom}}(x)$ is defined as follows\cite{XSS}:
$$ W_{\mathrm{hom}}(x) =
\begin{cases}
	0, & \text{if } x = 0,\\[4pt]
	2^{k-2}, & \text{if } x \notin 2^{k-1}\mathbb{Z}_{2^k},\\[4pt]
	2^{k-1}, & \text{if } x \in 2^{k-1}\mathbb{Z}_{2^k} \setminus \{0\}.
\end{cases} $$
Here $2^{k-1}\mathbb{Z}_{2^k}$ denotes the ideal generated by $2^{k-1}$ in $\mathbb{Z}_{2^k}$.
For vectors $\mathbf{x} = (x_1,\dots,x_n)$ and $\mathbf{y} = (y_1,\dots,y_n)$ in $\mathbb{Z}_{2^k}^n$, the \textbf{homogeneous distance} is defined as:
$$ d_{\mathrm{hom}}(\mathbf{x}, \mathbf{y}) = \sum_{i=1}^n W_{\mathrm{hom}}(y_i - x_i). $$
This function $d_{\mathrm{hom}}$ satisfies the axioms of a metric on $\mathbb{Z}_{2^k}^n$.

Following \cite{XSS}, we construct a weakly metric association scheme based on the ideal structure of the ring $\mathbb{Z}_{2^k}$. First, the alphabet is partitioned into four subsets: the zero set $Z = \{0\}$, the unit group $U = \mathbb{Z}_{2^k}^\times$, the principal ideal $S = \langle 2^{k-1}\rangle\setminus\{0\}$, and the set $V = \langle 2 \rangle \setminus \langle 2^{k-1} \rangle$. For a vector $\mathbf{x} = (x_1, \ldots, x_n) \in \mathbb{Z}_{2^k}^n$, define its partition statistic as $\pi(\mathbf{x}) = (\pi_Z(\mathbf{x}), \pi_U(\mathbf{x}), \pi_S(\mathbf{x}), \pi_V(\mathbf{x}))$, where $\pi_I(\mathbf{x}) = |\{j : x_j \in I\}|$, and it satisfies $\pi_Z + \pi_U + \pi_S + \pi_V = n$.
Let
\[D_\pi=\left\{\pi^{(i)}=\left(\pi_Z^{(i)},\pi_U^{(i)},\pi_S^{(i)},\pi_V^{(i)}\right):i=0,1,\ldots,s\right\},
\]
where
\[
s=
\begin{cases}
\displaystyle\binom{n+2}{2}-1,
& k=2,\\[6pt]
\displaystyle\binom{n+3}{3}-1,
& k\geq3.
\end{cases}
\]
Indeed, when $k=2$, the set $V$ is empty, so only three nonempty symbol classes occur.
An association scheme $(X, R)$ is defined on $\mathbb{Z}_{2^k}^n$, where $R = \{R_0, R_1, \ldots, R_s\}$ satisfies:
\[
\mathbf{x} R_i \mathbf{y} \iff \pi(\mathbf{y} - \mathbf{x}) = \pi^{(i)} \in D_{\pi}.
\]
Furthermore, a distance function is defined on this scheme:
\[
d(\mathbf{x},\mathbf{y})=\pi_U(\mathbf{y}-\mathbf{x})+\pi_V(\mathbf{y}-\mathbf{x})+2\pi_S(\mathbf{y}-\mathbf{x}).
\]
Since the homogeneous weight satisfies $W_{\mathrm{hom}}(x) = 0$ (if $x \in Z$), $2^{k-2}$ (if $x \in U \cup V$), and $2^{k-1}$ (if $x \in S \setminus \{0\}$), it is easy to verify that the homogeneous distance $d_{\mathrm{hom}}$ differs from $d$ only by a constant factor $2^{k-2}$, i.e., $d_{\mathrm{hom}}(\mathbf{x}, \mathbf{y}) = 2^{k-2} \cdot d(\mathbf{x}, \mathbf{y})$. Thus, this association scheme forms a weakly metric association scheme with respect to the homogeneous metric.

\subsection{$q$-ary Johnson scheme}
The binary Johnson scheme is a symmetric association scheme defined over the binary alphabet $F$ (i.e., \( F = \{0,1\} \)). Consider the set of all binary vectors of length \( n \), fix a weight \( w \) (with \( 0 < w < n \)), and let \( W_w(n,2) \) denote the set of all vectors with exactly \( w \) nonzero components. On this set, define the relations \( R_k = \{(x,y) \in W_w(n,2)^2 : d(x,y) = k\} \), where \( d(x,y) \) is the Hamming distance. Since the Hamming distance between two binary vectors of weight \( w \) must be even and is determined by the number of positions where both vectors have a 1, these relations form a symmetric association scheme, called the binary Johnson scheme \( J(w,n) \) \cite{D}. This scheme plays a significant role in coding theory and combinatorial design theory.

When the alphabet size \( q \geq 3 \) , the relations \( R_k \) based solely on Hamming distance do not in general make \( W_w(n,q) \) an association scheme, because the diversity of nonzero symbols requires a more refined description of similarity between vectors. Therefore, two parameters are introduced: the \textbf{number of equal nonzero coordinates} \( e(x,y) = |\{i : x_i = y_i \neq 0\}| \) and the \textbf{number of common nonzero coordinates} \( n(x,y) = |\{i : x_i \neq 0, y_i \neq 0\}| \). Based on these two parameters, the nonbinary Johnson scheme introduced in \cite{TAG} has relations
\[
R_{ij}=\left\{(x,y)\in W_w(n,q)^2:e(x,y)=w-i,\ n(x,y)=w-j\right\},
\]
where the relation indices range over
\[
\mathcal{D}_J=
\left\{
(i,j)\in\mathbb{N}^2:
0\leq j\leq\min\{w,n-w\},\
j\leq i\leq w
\right\}.
\]
For $(x,y)\in R_{ij}$, their Hamming distance is \(d_H(x,y)=i+j\). These relations form a symmetric association scheme, called the $q$-ary Johnson scheme and denoted by $J_q(w,n)$.

\subsection{The Lee association scheme}
The Lee association scheme \cite{S3}, denoted by $L(n, q)$, is defined as the extension of order $n$ of the ordinary $q$--gon scheme on $\Z_q$. This scheme has $s = \lfloor q/2 \rfloor$ classes, with relations defined by: $x R_k y \Leftrightarrow x-y = \pm k$.
For any vector $z$ in $\Z_q^n$, define its {\bf Lee composition} as $lc(z) = (c_0, c_1, \ldots, c_s)$,where \(c_i=\left|\left\{j\in\{1,\ldots,n\}:z_j\equiv\pm i\pmod q\right\}\right|,~0\leq i\leq s\), and $s=\lfloor q/2\rfloor$.

Note that the {\bf Lee weight} of $x$ is then $w_L(x)=\sum\limits_{i=0}^s i c_i.$ The {\bf Lee distance} between $x$ and $y$ is $d_L(x,y)=w_L(x-y).$
The Lee scheme on $\Z_q^n$ is then defined by the relations:
$$ x R_k y \Leftrightarrow lc(x - y) = k, $$
where $k$ ranges over all possible composition vectors. This scheme has $N = \binom{n+s}{s} - 1$ classes.
The Lee metric is constant on each class $R_k$, satisfying:
$$ x R_k y \Rightarrow d_L(x, y) = \|k\| = k_1 + 2k_2 + \cdots + s k_s. $$
Hence, $L(n, q)$ is weakly metric with respect to the Lee distance.
Note that the alternative definition based solely on distance values:
$ x R_k y \Leftrightarrow d_L(x, y) = k $ does not generally yield an association scheme, as verified in cases such as $n = 2, i = 2, j = 2, k = 4$.

By using properties of the inner product one can obtain a closed form for the generating function of the $p_k(i)$.\\
If we set:
$$ P_i(z) = \sum_{k} p_k(i) z_1^{k_1} \cdots z_s^{k_s}.$$
We obtain for $q$ odd:
$$ P_i(z) = \prod_{l=0}^{s} \left(1 + 2\sum_{m=1}^{s} z_m\cos\left(\frac{2\pi}{q}ml\right)\right)^{i_l}, $$
and for $q$ even:
$$ P_i(z) = \prod_{l=0}^{s} \left(1 + 2\sum_{m=1}^{s-1} z_m\cos\left(\frac{2\pi}{q}ml\right) + (-1)^{l} z_s\right)^{i_l}. $$
All these relations and the links with Krawtchouk polynomials show that the $p_k(i)$ are polynomials in $s$ variables $i_1, i_2, \ldots, i_s$ of total degree at most $|k|$.
Moreover, in a P and Q-polynomial scheme (i.e., where the $p_k(i)$ and $q_k(i)$ are polynomials in one variable) with sufficiently many classes, the $p_k(i)$ are rational numbers \cite[Theorem 7.11 p. 359]{BI}. However, the cosine terms do not affect multivariate polynomiality. Hence, the Lee
scheme (self-dual by \cite{B+}) is an $s$-polynomial WMAS in the sense of Section~2.4.

The dispersion function $\Pi_{q,n}(e)$ counts the number of distinct Lee compositions
in the Lee sphere of radius $e$ centered at the origin. Following \cite{S3}, we define the difference function:
$$ \delta_{q,n}(e) = \Pi_{q,n}(e) - \Pi_{q,n}(e-1), $$
This is equivalent to:
$$ \delta_{q,n}(e) = \left|\left\{ \mathbf{k} \in \mathbb{N}^s \,\middle|\, \sum_{i=1}^{s} k_i \leq n \text{ and } \sum_{i=1}^{s} i k_i = e \right\}\right|. $$
The function $\delta_{q,n}(e)$ enumerates the number of partitions of the integer $e$ into at most $n$ parts,
where each part has size in $[s]$.
Let $(x)_n = (1-x)(1-x^2)\cdots(1-x^n)$ and define the Gaussian binomial coefficient:
$$ \left[ \begin{array}{c} n \\ m \end{array} \right]_x = \frac{(x)_n}{(x)_m(x)_{n-m}}. $$
Then the generating function is given by:
$$ \sum_{e=0}^{\infty} \delta_{q,n}(e) x^e = \left[ \begin{array}{c} n+s \\ s \end{array} \right]_x = \left[ \begin{array}{c} n+s \\ n \end{array} \right]_x. $$
Consequently, for the dispersion function:
$$ \sum_{e=0}^{\infty} \Pi_{q,n}(e) x^e = \frac{1}{1-x} \left[ \begin{array}{c} n+s \\ n \end{array} \right]_x. $$

\subsection{Sum-rank metric}
Let $t$ be a positive integer and let $n_1,\ldots,n_t$ and $m_1,\ldots,m_t$ be positive integers with $m_j\geq n_j\geq1$ for all $1\leq j\leq t$. For positive integers $m\geq n$, let $\mathbb{F}_{q}^{n\times m}$ denote the vector space of all $n\times m$ matrices over the finite field $\mathbb{F}_{q}$. Denote by $\operatorname{rk}(M)$ the rank over $\mathbb{F}_{q}$ of a matrix $M\in\mathbb{F}_{q}^{n\times m}$.
We use the following compact notation for the direct sum of vector spaces of matrices:
$$ \operatorname{Mat}(\mathbf{n},\mathbf{m},\mathbb{F}_{q}):=\bigoplus_{i=1}^{t} \mathbb{F}_{q}^{n_{i}\times m_{i}}. $$
Equivalently, we may write this as the direct product:
$$ \mathbb{F}_{q}^{\mathbf{n}\times\mathbf{m}}:=\mathbb{F}_{q}^{n_{1}\times m_{1}} \times\cdots\times\mathbb{F}_{q}^{n_{t}\times m_{t}}, $$
where $\times$ stands for the direct product of vector spaces.

\begin{definition}(\cite{PRZ})
Let $X:=(X_{1},\ldots,X_{t})\in\operatorname{Mat}(\mathbf{n},\mathbf{m},\mathbb{F}_{q})$.
The \textbf{sum-rank weight} of $X$ is the quantity
$$ \operatorname{w}_{\operatorname{srk}}(X):=\sum_{i=1}^{t}\operatorname{rk}(X_{i}). $$
We also write $\operatorname{srk}(X):=\operatorname{w}_{\operatorname{srk}}(X)$.
\end{definition}

With this definition, we can endow the space \(\operatorname{Mat}(\mathbf{n},\mathbf{m},\mathbb{F}_{q})\) with a distance function, called the \textbf{sum-rank distance},
$$\operatorname{d}_{\operatorname{srk}}:\operatorname{Mat}(\mathbf{n},\mathbf{m},\mathbb{F}_{q})\times\operatorname{Mat}(\mathbf{n},\mathbf{m},\mathbb{F}_{q})\longrightarrow\mathbb{N} $$
defined by $\operatorname{d}_{\operatorname{srk}}(X,Y):=\operatorname{w}_{\operatorname{srk}}(X-Y)$.
Equivalently, the \textbf{sum-rank distance} between
$X,Y\in\mathbb{F}_{q}^{\mathbf{n}\times\mathbf{m}}$ is $\operatorname{srk}(X-Y)$.

It is easy to see that the sum-rank distance is indeed a metric on
$ \operatorname{Mat}(\mathbf{n},\mathbf{m},\mathbb{F}_{q}) $. In particular, if $t=1$ the sum-rank distance corresponds to the rank distance, and if all the $n_{i}$'s or all the $m_{i}$'s are equal to one, then the sum-rank distance coincides with the Hamming distance.

We can now give the definition of sum-rank metric codes \cite{AGKP}.
\begin{definition}
	A \textbf{sum-rank metric code} \(\mathcal{C}\) is a non-empty subset of
	$ \operatorname{Mat}(\mathbf{n},\mathbf{m},\mathbb{F}_{q}) $ endowed with the sum-rank distance. The \textbf{minimum sum-rank distance} of a code $\mathcal{C}$ with $|\mathcal{C}|\geq 2$ is defined by
	$$ \operatorname{srk}(\mathcal{C}):=\min\left\{\operatorname{srk}(X-Y)\colon X,Y \in\mathcal{C},\,X\neq Y\right\}. $$
	For an $\mathbb{F}_{q}$-linear subspace
	$ \mathcal{C}\subseteq\operatorname{Mat}(\mathbf{n},\mathbf{m},\mathbb{F}_{q}) $, we say that $ \mathcal{C}$ is an $\mathbb{F}_{q}$-linear sum-rank metric code.
\end{definition}

The sum-rank metric scheme is defined as the direct product of $t$ bilinear forms schemes \cite[280]{BCN}, \cite[306]{BI}. Each bilinear forms scheme is $P$-polynomial, which implies that its eigenvalues are given by univariate polynomials. For the direct product scheme, the eigenvalue
$P_{\mathbf{k}}(\mathbf{i})$ associated with a relation indexed by $\mathbf{k}=(k_1,\ldots,k_t)$
and a primitive eigenspace indexed by $\mathbf{i}=(i_1,\ldots,i_t)$ is given by
\[
P_{\mathbf{k}}(\mathbf{i})=\prod_{j=1}^{t}P_{k_j}^{(j)}(i_j).
\]
By the $P$-polynomial property of each factor, there exists a univariate polynomial
$v_{k_j}^{(j)}$ of degree $k_j$ such that \(P_{k_j}^{(j)}(i_j)=v_{k_j}^{(j)}\bigl(P_1^{(j)}(i_j)\bigr)\). Thus taking the spectral points
\(\mathbf{p}_{\mathbf{i}}=\bigl(P_1^{(1)}(i_1),\ldots,P_1^{(t)}(i_t)\bigr)\), we obtain
\(P_{\mathbf{k}}(\mathbf{i})=v_{\mathbf{k}}(\mathbf{p}_{\mathbf{i}})\), where
\(v_{\mathbf{k}}(x_1,\ldots,x_t)=\prod_{j=1}^{t}v_{k_j}^{(j)}(x_j)\). Since
\(\deg v_{\mathbf{k}}=\sum_{j=1}^{t}k_j=|\mathbf{k}|_1=d(\mathbf{k})\), the sum-rank metric scheme is a $t$-polynomial WMAS in the sense of Section~2.4.

\subsection{Mixed alphabet codes}
Mixed alphabet codes are coding structures defined over multiple alphabets of different sizes. Let $p_1, p_2, \ldots, p_n$ be positive integers $($not necessarily prime numbers$)$. Consider the Cartesian product space
$$ S = S_1 \times S_2 \times \cdots \times S_n, $$
where each component $S_i=\mathbb{Z}/p_i \mathbb{Z}$ denotes the ring of integers modulo $ p_i$. The elements of this space are $n$-tuples $s=(s_1, s_2, \ldots, s_n)$, where $s_i \in S_i$.

In this space, the Hamming distance between two elements $s, t \in S$ is defined as the number of coordinates in which they differ: $d(s, t) = |\{i \mid s_i \neq t_i\}|$.
The sphere of radius $e$ centered at $s \in S$ is defined as
$$ S_e(s) = \{t \in S \mid d(t, s) \leq e\}. $$
A subset $C \subseteq S$ is called a perfect $e$-code if for every $t \in S$, $|C \cap S_e(t)| = 1$.
In particular, when the alphabet sizes $p_i$ are not all equal, such codes are called {\bf mixed perfect codes}.

Let $\mathfrak{X}^{(k)} = (X^{(k)}, \{A_{i}^{(k)}\}_{i=0}^{d_k})$ be commutative association schemes of class $d_k$ for $k = 1, 2, \ldots, \ell$. The direct product of $\{\mathfrak{X}^{(k)}\}_{k=1}^{\ell}$ is the association scheme \cite{B+} defined by the Kronecker product of the adjacency matrices of $\mathfrak{X}^{(k)}$:
$$ A_{(n_1, n_2, \ldots, n_\ell)} := A_{n_1}^{(1)} \otimes A_{n_2}^{(2)} \otimes \cdots \otimes A_{n_\ell}^{(\ell)}, $$
for $(n_1, n_2, \ldots, n_\ell) \in D := \{0, 1, \ldots, d_1\} \times \{0, 1, \ldots, d_2\} \times \cdots \times \{0, 1, \ldots, d_\ell\}$. This association scheme is denoted by $\bigotimes_{k=1}^{\ell} \mathfrak{X}^{(k)}$.

In the context of mixed alphabet codes, let $F$ be a partition of the set $\{1,2,\ldots,n\}$ into subsets $A_{1},A_{2},\ldots,A_{k}$, with the size of each subset being $n_{i}=|A_{i}|$, where $i=1,2,\ldots,k$. Assume that all coordinates in each block $A_i$ have the same alphabet size $q_i$, that is, $p_j=q_i$ for every $j\in A_i$. Consider the direct product of Hamming schemes
$$  H(n_1,q_1) \otimes H(n_2,q_2) \otimes \cdots \otimes H(n_k,q_k), $$
where $H(n_i,q_i)$ denotes Hamming scheme corresponding to subset $A_i$, with parameters determined by the alphabet sizes within that subset. This product is weakly metric for the Hamming distance defined above.

Within this framework, the function $P(F,e)$ defined in \cite{MA} counts the number of integer vectors $(s_{1},s_{2},\ldots,s_{k})$ satisfying:
$$P(F,e)=\left|\left\{(s_{1},\ldots,s_{k})\in\mathbb{Z}^{k}\,\middle|\,\sum_{i=1}^{k}s_{i}\leq e,\quad 0\leq s_{i}\leq n_{i},\ i=1,\ldots,k\right\}\right|.$$
We observe that $P(F,e)$ is precisely the dispersion function $\Pi(e)$ for the said direct product.

For the direct product association scheme corresponding to mixed alphabet codes, \(\mathfrak{X}=\bigotimes_{j=1}^{k}H(n_j,q_j)\), the first eigenvalues are given by
\[
P_{\mathbf{a}}(\mathbf{b})=\prod_{j=1}^{k}P_{a_j}^{(j)}(b_j),
\]
where $P_{a_j}^{(j)}(b_j)$ is the Krawtchouk polynomial of the Hamming scheme $H(n_j,q_j)$.
It is known that when $a_j$ is fixed, the Krawtchouk polynomial $P_{a_j}^{(j)}(b_j)$ has degree $a_j$ in $b_j$. Consequently, $P_{\mathbf{a}}(\mathbf{b})$ is a multivariate polynomial in
$b_1,\ldots,b_k$ of total degree \(|\mathbf{a}|_1=\sum_{j=1}^{k}a_j\). Since $d(\mathbf{a})=|\mathbf{a}|_1$, the product scheme is a $k$-polynomial WMAS.

\subsection{The NRT metric}
Let $\mathcal{Q}$ be a finite alphabet of size $q$, viewed as an additive group modulo $q$. Consider  \cite{NR} the set $\mathcal{Q}^{r,n}$ of vectors of dimension $rn$ over $\mathcal{Q}$. Each vector $x$ can be written as a concatenation of $n$ blocks of length $r$:
$$ x = (x_{11},\ldots,x_{1r}; \ldots; x_{n1},\ldots,x_{nr}). $$
For any vector $x$, define its \textbf{shape} $\lambda = (\lambda_1, \ldots, \lambda_r)$, where $\lambda_i$ ($i=1,\ldots,r$) denotes the number of $r$-blocks in $x$ whose rightmost nonzero entry is in the $i$-th position counting from the beginning of the block.

For two vectors $x, y \in \mathcal{Q}^{r,n}$, if $\mathrm{shape}(x-y) = \lambda$, we write $x \sim_{\lambda} y$. The shape vector $\lambda = (\lambda_1, \ldots, \lambda_r)$ defines a partition of a nonnegative integer $N \leq n$ into $r$ parts. Let
$$ \Delta_{r,n} = \{ \lambda \in (\mathbb{Z}_{\geq 0})^r : \textstyle\sum_{i=1}^r \lambda_i \leq n \}, $$
denote the set of all such partitions. For brevity, we define the following notation:
$$ |\lambda| = \sum_{i=1}^r \lambda_i, \quad |\lambda|' = \sum_{i=1}^r i \lambda_i, \quad \lambda_0 = n - |\lambda|. $$
Let $x \in \mathcal{Q}^{r,n}$ be a vector of shape $\lambda$. By defining the weight function (norm) $\mathrm{w}(x) = |\lambda|'$, we can induce a metric $d_r(x, y) = \mathrm{w}(x-y)$ on $\mathcal{Q}^{r,n}$. We call the function $d_r$ the ordered weight.
The \textbf{NRT weight} is defined as:
$ \mathrm{w}_{\mathrm{NRT}}(x) = |\lambda|'. $
The corresponding \textbf{NRT distance} is given by:
$ d_{\mathrm{NRT}}(x, y) = \mathrm{w}_{\mathrm{NRT}}(x-y). $
The space $\mathcal{Q}^{r,n}$ equipped with this metric is called the \textbf{ordered Hamming space} or \textbf{NRT space}, denoted by $\overrightarrow{H}(q,n,r)$. When $r=1$, the NRT distance reduces to the classical Hamming distance.

An $(nr, M, d)$ ordered code $C \subseteq \mathcal{Q}^{r,n}$ is an arbitrary subset of $M$ vectors in $\mathcal{Q}^{r,n}$ such that the ordered distance between any two distinct vectors in $C$ is at least $d$. If $\mathcal{Q}$ is a finite field and $C$ is a linear code of dimension $k$, we refer to it as an $[nr, k, d]$ code.
An association scheme can be defined on the NRT space $\overrightarrow{H}(q,n,r)$, where the relations are given by shape partitions. Specifically, for each shape $\lambda \in \Delta_{r,n}$, define the relation:
$$ R_\lambda = \{ (x, y) \in \mathcal{Q}^{r,n} \times \mathcal{Q}^{r,n} : \mathrm{shape}(x-y) = \lambda \}. $$
This association scheme is called the \textbf{ordered Hamming scheme}.

For the NRT scheme (ordered Hamming scheme) \(\overrightarrow{\mathcal{H}}(q,n,r)\), the first eigenvalues are given by the multivariate Krawtchouk polynomials: for any relation index
$\mathbf{f}=(f_1,\ldots,f_r)\in\Delta_{r,n}$ and any primitive eigenspace index $\mathbf{e}=(e_1,\ldots,e_r)\in\Delta_{r,n}$, we have $P_{\mathbf{f}}(\mathbf{e})
=K_{\mathbf{f}}(\mathbf{e})$. When \(\mathbf{f}\) is fixed, \(K_{\mathbf{f}}(\mathbf{e})\) is a multivariate polynomial in the variables \(e_1, \dots, e_r\) of degree \(|\mathbf{f}| = \sum_{i=1}^{r} f_i\). Since the NRT distance associated with $\mathbf{f}$ is
$d(\mathbf{f})=|\mathbf{f}|'$, we have
\[
\deg K_{\mathbf{f}}=|\mathbf{f}|=\sum_{i=1}^{r}f_i\leq\sum_{i=1}^{r}if_i=d(\mathbf{f}).
\]
The relation indices $\mathbf{f}$ and the primitive eigenspace indices $\mathbf{e}$ both range over
$\Delta_{r,n}\subseteq\{0,1,\ldots,n\}^r$. Hence, the NRT scheme is an $r$-polynomial WMAS.

\section{A Lloyd theorem for perfect codes}
\subsection{Perfect Codes}
Let \(\mathfrak X=(X,\{R_\alpha\}_{\alpha\in \mathcal D})\) be a finite symmetric commutative association scheme, where \(\mathcal D\subseteq \mathbb N^m\) indexes the nonempty relations of \(\mathfrak X\).
Let \(A_\alpha\) be the adjacency matrix of \(R_\alpha\). Assume that \(\mathfrak X\) is an \(m\)-variate \(P\)-polynomial association scheme with respect to a monomial order \(\leq\) on \(\mathcal D\), then there exist adjacency matrices \(A_{\epsilon_1},\ldots,A_{\epsilon_m}\) and polynomials \(v_\alpha(x_1,\ldots,x_m)\) such that
\( A_\alpha = v_\alpha(A_{\epsilon_1},\ldots,A_{\epsilon_m}), \alpha\in \mathcal D \).
Assume further that the relations are induced by an \(m\)-distance \(d_m:X\times X\to \Delta\).

In this subsection, we identify \(\mathcal D\) with \(\Delta\) through
\((x,y)\in R_\alpha \iff d_m(x,y)=\alpha\).
Define $\widetilde d(x,y)=|d_m(x,y)|_1$. Assume that $\mathfrak X$ is weakly metric
with respect to $\widetilde d$. By Proposition~\ref{weakly metric}, this property is guaranteed when the monomial order is $L^1$ compatible with $\mathcal D$.
For \(e\ge 0\), define
\[
T_e=\{\alpha\in \mathcal D:|\alpha|_1\le e\},~\Pi(e)=|T_e|,~B_e=\sum_{\alpha\in T_e}A_\alpha.
\]
Let $J=\{0,1,\ldots,s\}$ be the index set of the primitive idempotents
$\{E_j\}_{j\in J}$ of the Bose--Mesner algebra of $\mathfrak X$ with
$E_0=\frac{1}{|X|}J_X$. Using the spectral points $p_j$ introduced in Section~2.4, i.e. \(p_j=\bigl(P_{\epsilon_1}(j),\ldots,P_{\epsilon_m}(j)\bigr)\in\mathbb C^m,~j\in J\), define the multivariate Lloyd polynomial of radius $e$ by
\[
\Psi_e(x_1,\ldots,x_m)
=\sum_{\alpha\in T_e}v_\alpha(x_1,\ldots,x_m).
\]

{\thm \label{Lloyd}
If \(C\subseteq X\) is an \(e\)-perfect code with respect to \(\widetilde d\) and characteristic vector \(\chi_C\), then there exist at least \(\Pi(e)-1\) indices \(j\in J\setminus\{0\}\) s.t. \( E_j\chi_C\ne 0~ \text{and}~ \Psi_e(p_j)=0. \)}
\begin{proof}
Since \(C\) is \(e\)-perfect with respect to \(\widetilde d\), every \(x\in X\) has a unique codeword \(c\in C\) satisfying \(\widetilde d(x,c)\le e\). For each \(x\in X\), we have \((B_e\chi_C)_x = \sum_{\alpha\in T_e}(A_\alpha\chi_C)_x\). \\
Moreover
\[
(A_\alpha\chi_C)_x = |\{c\in C:(x,c)\in R_\alpha\}| = |\{c\in C:d_m(x,c)=\alpha\}|.
\]
Therefore
\[
(B_e\chi_C)_x = |\{c\in C:\widetilde d(x,c)\le e\}| = 1.
\]
Hence \(B_e\chi_C=\mathbf 1_X\), where \(\mathbf 1_X\) denotes all-one column vector and length \(|X|\). Since \( A_\alpha = v_\alpha(A_{\epsilon_1},\ldots,A_{\epsilon_m})~ \text{and}~ A_{\epsilon_i}E_j=P_{\epsilon_i}(j)E_j\), we get
\[
A_\alpha E_j = v_\alpha\bigl(P_{\epsilon_1}(j),\ldots,P_{\epsilon_m}(j)\bigr)E_j = v_\alpha(p_j)E_j.
\]
Thus
\[
B_eE_j = \sum_{\alpha\in T_e}A_\alpha E_j = \sum_{\alpha\in T_e}v_\alpha(p_j)E_j = \Psi_e(p_j)E_j.
\]
Applying \(E_j\) to \(B_e\chi_C=\mathbf 1_X\), and using the fact that \(B_e\) and \(E_j\) both lie in the Bose-Mesner algebra and hence commute, gives \( B_eE_j\chi_C = E_j\mathbf 1_X\). If \(j\ne 0\), then \(E_j\mathbf 1_X=0\), hence \(B_eE_j\chi_C=0\). Together with \(B_eE_j=\Psi_e(p_j)E_j\), this gives \(\Psi_e(p_j)E_j\chi_C=0\). Therefore, if \(j\ne 0\) and \(E_j\chi_C\ne 0\), then \(\Psi_e(p_j)=0\).

Consider \(\mathcal W_e=\{A_\alpha\chi_C:\alpha\in T_e\}\). Since \(R_\alpha\) is non-empty, it has positive constant valency. Hence for every \(c\in C\), there exists \(x\in X\) such that \((c,x)\in R_\alpha\). Because the scheme is symmetric, we also have \((x,c)\in R_\alpha\). So \( (A_\alpha\chi_C)_x\neq0\), then \(A_\alpha\chi_C\neq0\).\\
Let \(\alpha,\beta\in T_e\) with \(\alpha\ne\beta\). Suppose that some \(x\in X\) lies in both supports \(\operatorname{supp}(A_\alpha\chi_C)\) and \(\operatorname{supp}(A_\beta\chi_C)\), then there exist \(c,c'\in C\) such that
\(d_m(x,c)=\alpha,~ d_m(x,c')=\beta\).
Since \(\alpha,\beta\in T_e\), we have
\(\widetilde d(x,c)\le e,~ \widetilde d(x,c')\le e\).
By the uniqueness condition for an \(e\)-perfect code, we get \(c=c'\), which forces \(\alpha=\beta,\) a contradiction. Therefore,
\[
\operatorname{supp}(A_\alpha\chi_C) \cap \operatorname{supp}(A_\beta\chi_C) = \varnothing,~ \alpha\ne\beta.
\]
Thus the vectors \(A_\alpha\chi_C\), \(\alpha\in T_e\), are non-zero and pairwise orthogonal. In particular, they are linearly independent, and so
\[
\dim \operatorname{Span}\{A_\alpha\chi_C:\alpha\in T_e\} = |T_e| = \Pi(e).
\]
On the other hand, the primitive idempotent decomposition gives
\(A_\alpha = \sum_{j\in J}P_\alpha(j)E_j\).
Hence \(A_\alpha\chi_C = \sum_{j\in J}P_\alpha(j)E_j\chi_C\). Therefore,
\[
\operatorname{Span}\{A_\alpha\chi_C:\alpha\in T_e\} \subseteq \operatorname{Span}\{E_j\chi_C:j\in J,\ E_j\chi_C\ne 0\}.
\]
Define \(r(C)= \left|\{j\in J\setminus\{0\}:E_j\chi_C\ne 0\}\right|\).
Comparing dimensions yields \(\Pi(e)\le 1+r(C)\), and hence \(r(C)\ge \Pi(e)-1\).
Therefore, there exist at least \(\Pi(e)-1\) indices \(j\in J\setminus\{0\}\) such that \(E_j\chi_C\neq0\) and \(\Psi_e(p_j)=0\).
\end{proof}

The following example illustrates Theorem~\ref{Lloyd} in a genuinely multivariate setting. We use a construction of $1$-perfect mixed codes due to Etzion \cite[Theorem~7.1]{E}. The same construction is also restated in \cite[Theorem~1]{Tuvi}.

{\ex
Let $\mathbb{F}_8=\mathbb{F}_2^3$, and consider the subspaces
\[
\mathcal{T}_1=\{000,110,011,101\},
\]
\[
\mathcal{T}_2=\{000,100\},~\mathcal{T}_3=\{000,010\},~\mathcal{T}_4=\{000,001\},~\mathcal{T}_5=\{000,111\}.
\]
The sets \(\mathcal{T}_i\setminus\{0\},~1\leq i\leq5\) form a partition of $\mathbb{F}_8^\ast$. By \cite[Theorem~7.1]{E},
\[
C=\left\{(c_1,\ldots,c_5)\in\prod_{i=1}^5\mathcal{T}_i:\sum_{i=1}^5c_i=0\right\}
\]
is a $1$-perfect mixed code.
Set \(X=\prod_{i=1}^5\mathcal{T}_i\), the association scheme on $X$ is \(K_4\otimes K_2^4\) with relations indexed by \(\mathcal D=\{0,1\}^5\). If \(d_m(x,y)=\alpha=(\alpha_1,\ldots,\alpha_5)\), where $\alpha_i=1$ if and only if $x_i\neq y_i$, then
\[
\widetilde d(x,y)=|\alpha|_1=|\{i:x_i\neq y_i\}|,
\]
which is exactly the mixed Hamming distance. So $C$ is $1$-perfect in Theorem~\ref{Lloyd}.
For $e=1$, \(T_1=\{0,\epsilon_1,\ldots,\epsilon_5\}\), and then \(\Pi(1)=6\).
Write
\[
C=\ker\Phi,~\Phi(c_1,\ldots,c_5)=c_1+\cdots+c_5.
\]
For $a\in\mathbb{F}_2^3$,
define
\[
\psi_a(c_1,\ldots,c_5)=(-1)^{a\cdot(c_1+\cdots+c_5)}.
\]
Let \(C^\circ=\{\psi\in\widehat X:\psi(c)=1\ \text{for all }c\in C\}\) be the annihilator of $C$. Since $C=\ker\Phi$, the eight characters $\psi_a~(a\in\mathbb{F}_2^3)$ form $C^\circ$.
Hence, by Fourier inversion,
\[
\chi_C=\frac{|C|}{|X|}\sum_{\psi\in C^\circ}\psi=\frac18\sum_{a\in\mathbb{F}_2^3}\psi_a.
\]
The primitive eigenspaces of $K_4\otimes K_2^4$ are indexed by
\(\gamma=(\gamma_1,\ldots,\gamma_5)\in\{0,1\}^5\),
where $\gamma_i=0$ or $1$ according as the restriction of the character to $\mathcal{T}_i$ is trivial or nontrivial. For
$a=(a_1,a_2,a_3)$,
\[
\gamma_2=a_1,~\gamma_3=a_2,~\gamma_4=a_3,~\gamma_5=a_1+a_2+a_3 \pmod 2,
\]
while
\(\mathcal{T}_1^\perp=\{000,111\}\). Thus the seven nontrivial characters correspond to
\[
\begin{array}{c|c}
a & \gamma\\
\hline
100 &(1,1,0,0,1)\\
010 &(1,0,1,0,1)\\
001 &(1,0,0,1,1)\\
110 &(1,1,1,0,0)\\
101 &(1,1,0,1,0)\\
011 &(1,0,1,1,0)\\
111 &(0,1,1,1,1).
\end{array}
\]
Since each $\psi_a$ occurs with nonzero coefficient in the Fourier expansion of $\chi_C$ and belongs to the corresponding primitive eigenspace $V_\gamma$, we have
\(E_\gamma\chi_C\neq0\). Hence
\[
|S^*(C)|=7\geq5=\Pi(1)-1.
\]
Finally, for this product scheme, \(v_0=1,~v_{\epsilon_i}(x_1,\ldots,x_5)=x_i\),
so
\[
\Psi_1(x_1,\ldots,x_5)=1+x_1+x_2+x_3+x_4+x_5.
\]
Since the nontrivial adjacency matrix of $K_q$ has eigenvalues
$q-1$ and $-1$, we obtain
\[
p_\gamma
=
\bigl(
3-4\gamma_1,\,
1-2\gamma_2,\,
1-2\gamma_3,\,
1-2\gamma_4,\,
1-2\gamma_5
\bigr).
\]
Therefore
\[
\Psi_1(p_\gamma)
=
8-4\gamma_1
-2(\gamma_2+\gamma_3+\gamma_4+\gamma_5).
\]
We can obtain \(\Psi_1(p_\gamma)=0\) for all the seven indices above. Thus
\[
|S^*(C)|=7\geq 5=\Pi(1)-1,~\text{and}~\Psi_e(p_j)=0,~j\in S^*(C).
\]
as asserted by Theorem~\ref{Lloyd}.
}

\section{Outer distribution matrix}
Let \(C\subseteq X\) be a code. We use the outer distribution matrix introduced in Section~2.5,
\[
B(C)=\bigl(B_\alpha\bigr)_{\alpha\in D},~ B_\alpha=A_\alpha\chi_C.
\]
For each \(x\in X\),
\[
(B_\alpha)_x = |\{c\in C:(x,c)\in R_\alpha\}|.
\]
Define the nontrivial external support of \(C\) by
\[
S^*(C) = \{j\in J\setminus\{0\}:E_j\chi_C\neq 0\}.
\]
Its cardinality \( r(C)=|S^*(C)|\) is called the \textbf{external distance} of \(C\).

For \(f\in\mathbb C[x_1,\ldots,x_m]\), set \(M_f=f(A_{\epsilon_1},\ldots,A_{\epsilon_m})\).
Since \(M_f\) belongs to the Bose-Mesner algebra, and since
\(\{A_\alpha:\alpha\in\mathcal D\}\) is a basis of this algebra, there
exist unique scalars \(c_\alpha(f)\), \(\alpha\in\mathcal D\), such that
\(M_f=\sum_{\alpha\in\mathcal D}c_\alpha(f)A_\alpha\).

{\thm \label{Col}
Let \(f\in\mathbb C[x_1,\ldots,x_m]\) satisfy \(f(p_j)=0, j\in S^*(C)\), then
\( \sum_{\alpha\in\mathcal D}c_\alpha(f)B_\alpha = \frac{|C|}{|X|}f(p_0)\mathbf 1_X \), where \(\mathbf 1_X\) denotes the all-one column vector of length \(|X|\). Furthermore, if \(M_f\neq0\) and \(\gamma=\max_{\leq}\{\alpha\in\mathcal D:c_\alpha(f)\neq0\}\), then
\(B_\gamma \in \operatorname{Span} \left( \mathbf 1_X,\{B_\alpha:\alpha<\gamma\} \right). \) }
\begin{proof}
Since \(A_{\epsilon_i}E_j=P_{\epsilon_i}(j)E_j\), we obtain
\[
M_fE_j=f(A_{\epsilon_1},\ldots,A_{\epsilon_m})E_j=f\bigl(P_{\epsilon_1}(j),\ldots,P_{\epsilon_m}(j)\bigr)E_j=f(p_j)E_j.
\]
By the primitive idempotent decomposition \(I=\sum_{j \in J}E_j\), we have
\(\chi_C = \sum_{j\in J}E_j\chi_C \). Therefore
\[
M_f\chi_C = \sum_{j\in J}M_fE_j\chi_C = \sum_{j\in J}f(p_j)E_j\chi_C.
\]
For \(j\neq0\), if \(j\in S^*(C)\), then \(f(p_j)=0\). If \(j\notin S^*(C)\), then \(E_j\chi_C=0\). Hence all nontrivial terms vanish, and only the term corresponding to \(j=0\) remains. Thus
\[
M_f\chi_C = f(p_0)E_0\chi_C.
\]
Since \(E_0=\frac{1}{|X|}J_X\), we get
\[
M_f\chi_C = f(p_0)\frac{1}{|X|}J_X\chi_C = f(p_0)\frac{|C|}{|X|}\mathbf 1_X.
\]
On the other hand, by the definition of the coefficients \(c_\alpha(f)\),
\[
M_f\chi_C = \sum_{\alpha\in\mathcal D}c_\alpha(f)A_\alpha\chi_C = \sum_{\alpha\in\mathcal D}c_\alpha(f)B_\alpha.
\]
Thus
\[
\sum_{\alpha\in\mathcal D}c_\alpha(f)B_\alpha = f(p_0)\frac{|C|}{|X|}\mathbf 1_X.
\]
Now assume \(M_f\neq0\), then the set \(\{\alpha\in\mathcal D:c_\alpha(f)\neq 0\}\) is non-empty. Since \(\mathcal D\) is finite and \(\leq\) is a monomial order, it has a maximum element \(\gamma=\max_{\leq}\{\alpha\in\mathcal D:c_\alpha(f)\neq 0\}\).
Thus \(c_\gamma(f)\neq0\), and \(c_\alpha(f)=0\) for every \(\alpha>\gamma\). The column relation becomes
\[
B_\gamma = \frac{|C|}{|X|c_\gamma(f)}f(p_0)\mathbf 1_X - \sum_{\alpha<\gamma}\frac{c_\alpha(f)}{c_\gamma(f)}B_\alpha.
\]
Therefore
\[
B_\gamma \in \operatorname{Span} \left( \mathbf 1_X,\{B_\alpha:\alpha<\gamma\} \right).
\]
\end{proof}

{\ex \label{example2}
Consider the $1$-perfect Lee code in \(\mathbb Z_5^2\) appearing in \cite[Example IV.1]{Kim}.
\[
C=\langle(3,1)\rangle=\{(0,0),(3,1),(1,2),(4,3),(2,4)\}\subseteq \mathbb{Z}_5^2.
\]
The Lee composition classes are indexed by
\((0,0),(1,0),(0,1),(2,0),(1,1),(0,2)\).
Let \(\omega=e^{2\pi i/5}\). For $u=(u_1,u_2)\in\mathbb{Z}_5^2$, define the addtive character
\(\psi_u(x_1,x_2)=\omega^{u_1x_1+u_2x_2}\). Since \(C=\{t(3,1):t\in\mathbb{Z}_5\}\), the Fourier coefficient of $\chi_C$ corresponding to $\psi_u$ is
\[
\widehat{\chi_C}(u)=\frac{1}{25}\sum_{t=0}^{4}\omega^{-t(3u_1+u_2)}.
\]
Hence
\[
\widehat{\chi_C}(u)\neq 0~ \Longleftrightarrow~ 3u_1+u_2\equiv 0\pmod{5}.
\]
Therefore
\[
C^\perp=\{(0,0),(1,2),(2,4),(3,1),(4,3)\}.
\]
All nonzero elements of $C^\perp$ have Lee composition $(1,1)$. Since the primitive eigenspaces of the Lee translation scheme are indexed by the Lee compositions of the characters, it follows that \(S^*(C)=\{(1,1)\}\). Choose $u=(1,2)$ as a representative character of Lee composition $(1,1)$. For each relation $\alpha$,
\[
P_\alpha((1,1))=\sum_{\operatorname{lc}(z)=\alpha}\psi_u(z)=\sum_{\operatorname{lc}(z)=\alpha}\omega^{z_1+2z_2}.
\]
For relation $(1,0)$, the possible values of $z$ are \((\pm1,0),~(0,\pm1)\), and hence
\[
P_{(1,0)}((1,1))=\omega+\omega^4+\omega^2+\omega^3=-1.
\]
For relation $(0,1)$, the possible values of $z$ are \((\pm2,0),~(0,\pm2)\), and hence
\[
P_{(0,1)}((1,1))=\omega^2+\omega^3+\omega^4+\omega=-1.
\]
Thus the corresponding spectral point is
\[
p_{(1,1)}=\bigl(P_{(1,0)}((1,1)),P_{(0,1)}((1,1))\bigr)=(-1,-1).
\]
Similarly, we have
\[
P_{(2,0)}((1,1))=-1;~ P_{(0,2)}((1,1))=-1;~ P_{(1,1)}((1,1))=3.
\]
Since \(k_{(1,0)}=k_{(0,1)}=4\), the trivial spectral point is \(p_0=(4,4)\).

We apply Theorem~\ref{Col} at $p_{(1,1)}$. Taking \(f_1(x_1,x_2)=x_1+1\), we have
\(M_{f_1}=A_{(1,0)}+A_{(0,0)},~f_1(p_0)=5\). Since $|C|/|X|=1/5$, \(B_{(0,0)}+B_{(1,0)}=\mathbf{1}_X\), and thus
\[
B_{(1,0)}=\mathbf{1}_X-B_{(0,0)}.
\]
Taking \(f_2(x_1,x_2)=x_1-x_2\), we have \(M_{f_2}=A_{(1,0)}-A_{(0,1)},~f_2(p_{(1,1)})=f_2(p_0)=0\), thus
\[
B_{(0,1)}=B_{(1,0)}.
\]
Next, since \(P_{(2,0)}((1,1))=P_{(0,2)}((1,1))=P_{(1,0)}((1,1))=-1\) and
\(k_{(2,0)}=k_{(0,2)}=k_{(1,0)}=4\), the polynomials \(f_3=v_{(2,0)}-x_1,~f_4=v_{(0,2)}-x_1\) vanish at both $p_{(1,1)}$ and $p_0$, which yields
\[
B_{(2,0)}=B_{(0,2)}=B_{(1,0)}.
\]
Finally \(P_{(1,1)}((1,1))=3\), then \(f_5=v_{(1,1)}-x_1-4\) vanishes at $p_{(1,1)}$. Since \(M_{f_5}=A_{(1,1)}-A_{(1,0)}-4A_{(0,0)}\) and
\(f_5(p_0)=k_{(1,1)}-k_{(1,0)}-4=8-4-4=0\), which gives
\[
B_{(1,1)}=4B_{(0,0)}+B_{(1,0)}.
\]
Here the maximal relation index occurring in $M_{f_5}$ is \(\gamma=(1,1)\), which also illustrates the second assertion of Theorem~\ref{Col}.

Consequently,
\[
B_{(0,1)}=B_{(2,0)}=B_{(0,2)}=B_{(1,0)}=\mathbf{1}_X-B_{(0,0)}
\]
and
\[
B_{(1,1)}=\mathbf{1}_X+3B_{(0,0)}.
\]
Since \(B_{(0,0)}=\chi_C\), if the rows are ordered with the elements of $C$ first and those of $X\setminus C$ afterwards, and the columns are ordered as
\[
(0,0),(1,0),(0,1),(2,0),(1,1),(0,2),
\]
then writing $\mathbf{1}_t$ and $\mathbf{0}_t$ for the all-one and all-zero column vectors of length $t$ respectively, the outer distribution matrix is
\[
B=
\begin{pmatrix}
\mathbf{1}_5
&
\mathbf{0}_5
&
\mathbf{0}_5
&
\mathbf{0}_5
&
4\mathbf{1}_5
&
\mathbf{0}_5
\\
\mathbf{0}_{20}
&
\mathbf{1}_{20}
&
\mathbf{1}_{20}
&
\mathbf{1}_{20}
&
\mathbf{1}_{20}
&
\mathbf{1}_{20}
\end{pmatrix}.
\]
The two displayed lines are block rows corresponding to $C$ and $X\setminus C$ respectively. Thus
\[
\operatorname{Col}(B)=\operatorname{Span}\{\mathbf{1}_X,B_{(0,0)}\},~\operatorname{rank}B=2.
\]
}

{\thm Let \(U\subseteq \mathcal D\). Suppose that \(\mathbf 1_X\in \operatorname{Span}\{B_\alpha:\alpha\in U\}\), then for every \(x\in X\), there exists \(\alpha\in U\) such that \((x,c)\in R_\alpha\). In particular, if the relations are induced by an \(m\)-distance \(d_m\) and \(U=T_R=\{\alpha\in\mathcal D:|\alpha|_1\le R\}\), then the covering radius of $C$ with respect to $\widetilde d$ satisfies \(\rho(C)\le R\).}
\begin{proof}
Since \(\mathbf 1_X\in\operatorname{Span}\{B_\alpha:\alpha\in U\}\), there exist scalars \(a_\alpha\),~\(\alpha\in U\), such that \(\mathbf 1_X=\sum_{\alpha\in U}a_\alpha B_\alpha\).
Suppose to the contrary that there exists \(x\in X\) such that \((B_\alpha)_x=0, \alpha\in U\). Taking the \(x\)-coordinate gives
\[
1=(\mathbf 1_X)_x=\sum_{\alpha\in U}a_\alpha (B_\alpha)_x=0,
\]
a contradiction. Thus for every \(x\in X\), there exists \(\alpha\in U\) such that \((B_\alpha)_x>0\). By the definition of \(B_\alpha\), there exists \(c\in C\) such that \((x,c)\in R_\alpha\). If the relations are induced by $d_m$ and $U=T_R$, then for every
$x\in X$ there exists $c\in C$ such that
\[
\widetilde d(x,c)=|d_m(x,c)|_1=|\alpha|_1\le R.
\]
Hence
\[
\rho(C)=\max_{x\in X}\min_{c\in C}\widetilde d(x,c)\le R.
\]
\end{proof}

{\ex
For the $1$-perfect Lee code \(C=\langle(3,1)\rangle\subseteq\mathbb{Z}_5^2\) considered in Example~\ref{example2}, we have
\[
B_{(0,0)}+B_{(1,0)}=\mathbf{1}_X.
\]
Since $(0,0),(1,0)\in T_1$, it follows that
\[
\mathbf{1}_X\in\operatorname{Span}\{B_\alpha:\alpha\in T_1\}.
\]
Hence the theorem gives \(\rho(C)\le1\). Since $C$ is a $1$-perfect Lee code, its covering radius is \(\rho(C)=1\), which is consistent with the conclusion of the theorem.\\
}
\section{Clique/coclique bound}
We recall Delsarte's clique-coclique bound \cite[Theorem 3.9]{D2} in the notation of the present paper.
{\lem\label{Delsarte's clique-coclique bound}
Let \(\mathfrak X=(X,\{R_\alpha\}_{\alpha\in\mathcal D})\) be a finite commutative association scheme. Let \(M\subseteq\mathcal D\) with \(\mathbf0\in M\), and set
\(M^*=M\setminus\{\mathbf0\}, \overline M=\mathcal D\setminus M^*\). If \(Y\subseteq X\) is an \(M\)-clique and \(Z\subseteq X\) is an \(\overline M\)-clique, then
\(|Y||Z|\le |X|\).
}

In our setting this translates into the following result.
{\coro
Assume that the relations are induced by an \(m\)-distance \(d_m\), and let
\(\widetilde d(x,y)=|d_m(x,y)|_1\). If \(C\subseteq X\) has minimum \(\widetilde d\)-distance at least \(d\), and \(A\subseteq X\) has \(\widetilde d\)-diameter at most \(d-1\), then \(|C||A|\le |X|\).

}
\begin{proof}
Let
\[
M_d=\{\mathbf 0\}\cup\{\alpha\in\mathcal D:|\alpha|_1\ge d\}.
\]
Since \(C\) has minimum \(\widetilde d\)-distance at least \(d\), it is an \(M_d\)-clique.
Moreover,
\[
\overline M_d=\mathcal D\setminus(M_d\setminus\{\mathbf 0\})=\{\alpha\in\mathcal D:|\alpha|_1\le d-1\}.
\]
Since \(A\) has \(\widetilde d\)-diameter at most \(d-1\), it is an
\(\overline M_d\)-clique. The result follows from Lemma \ref{Delsarte's clique-coclique bound}.
\end{proof}

We are now ready for the main result of this subsection, which is new, even in the single variable case.
{\thm \label{clique-colique bound}
Let $\mathfrak{X}=(X,\{R_\alpha\}_{\alpha\in \mathcal D})$ be a finite symmetric commutative $m$-variate $P$-polynomial association scheme. Suppose that \(C\subseteq X\) is an \(M\)-clique and \(A\subseteq X\) is an \(\overline{M}\)-clique. Let \(a=(a_\alpha)_{\alpha\in \mathcal D}\) be the inner distribution of \(A\), and define
\[
L_A(x_1,\ldots,x_m)=\sum_{\alpha\in \mathcal D}\frac{a_\alpha}{k_\alpha}v_\alpha(x_1,\ldots,x_m),
\]
where \(k_\alpha\) is the valency of \(R_\alpha\) and \(A_\alpha=v_\alpha(A_{\epsilon_1},\ldots,A_{\epsilon_m}).\)
Then equality holds in the clique--coclique bound if and only if
\(L_A(p_j)=0,~\forall j\in S^*(C)\).
}

\begin{proof}
Let \(a=(a_\alpha)_{\alpha\in \mathcal D}\) and \(c=(c_\alpha)_{\alpha\in \mathcal D}\) be the inner distributions of \(A\) and \(C\) respectively. By Delsarte's equality condition for the clique-coclique bound,
\[
|A||C|=|X|~\Longleftrightarrow~(aQ)_j(cQ)_j=0,~ \forall j\in J\setminus\{0\}.
\tag{1}
\]
By the relation between the MacWilliams transform of the inner distribution and the primitive idempotents,
\[
(cQ)_j=\frac{|X|}{|C|}\chi_C^{T}E_j\chi_C=\frac{|X|}{|C|}\|E_j\chi_C\|^2.
\]
Hence for \(j\in J\setminus\{0\}\),
\[
(cQ)_j>0 ~\Longleftrightarrow ~E_j\chi_C\neq0 ~\Longleftrightarrow ~j\in S^*(C).
\tag{2}
\]

On the other hand, the first and second eigenmatrices satisfy
\(\mu_jP_\alpha(j)=k_\alpha Q_j(\alpha)\), where \(\mu_j=\operatorname{rank}(E_j)\). Since \(A_\alpha=v_\alpha(A_{\epsilon_1},\ldots,A_{\epsilon_m})\), we have
\(P_\alpha(j)=v_\alpha(p_j)\). Thus,
\[
(aQ)_j=\sum_{\alpha\in \mathcal D}a_\alpha Q_j(\alpha)\\=\mu_j\sum_{\alpha\in \mathcal D}\frac{a_\alpha}{k_\alpha}P_\alpha(j)\\=\mu_j\sum_{\alpha\in \mathcal D}\frac{a_\alpha}{k_\alpha}v_\alpha(p_j)\\=\mu_jL_A(p_j).
\tag{3}
\]
Since \(\mu_j>0\), it follows that
\[
(aQ)_j=0 ~\Longleftrightarrow ~L_A(p_j)=0.
\tag{4}
\]	
Combining \((1)\), \((2)\), and \((4)\), we conclude that
\[
|C||A|=|X| ~\Longleftrightarrow ~L_A(p_j)=0, ~\forall j\in S^*(C).
\]
\end{proof}

The following example illustrates the equality criterion in Theorem~\ref{clique-colique bound}. We use a classical $1$-perfect Lee code which has already appeared above, and verify the spectral condition explicitly.

{\ex
We continue with the Lee scheme $L(2,5)$ and the $1$-perfect Lee code \(C=\langle(3,1)\rangle\subseteq \mathbb{Z}_5^2\) from Example~\ref{example2}. Every nonzero element of $C$ has Lee composition $(1,1)$, so $C$ is an $M$-clique with
\(M=\{(0,0),(1,1)\}\).
Let
\[
A=B_L((0,0),1)=\{(0,0),(1,0),(4,0),(0,1),(0,4)\}.
\]
The relation $(1,1)$ does not occur between any two points of $A$, hence $A$ is an $\overline{M}$-clique, where \(\overline{M}=D\setminus\bigl(M\setminus\{(0,0)\}\bigr)\).
Moreover, \(|C|=|A|=5\) and \(|X|=25\), so \(|C||A|=|X|\).
	
The valencies of the six Lee composition classes are \((1,4,4,4,8,4)\). Direct counting of ordered pairs in $A\times A$ gives the inner distribution \(a(A)=\left(1,\frac85,\frac45,\frac85,0,0\right)\). Therefore,
\[
L_A(x_1,x_2)=1+\frac25v_{(1,0)}(x_1,x_2)+\frac15v_{(0,1)}(x_1,x_2)+\frac25v_{(2,0)}(x_1,x_2).
\]	
From the previous example, \(S^*(C)=\{(1,1)\},~p_{(1,1)}=(-1,-1)\), and
\(P_{(1,0)}((1,1))=P_{(0,1)}((1,1))=P_{(2,0)}((1,1))=-1\).
Since \(v_\alpha(p_j)=P_\alpha(j)\), we obtain
\[
L_A(p_{(1,1)})=1+\frac{2}{5}P_{(1,0)}((1,1))+\frac{1}{5}P_{(0,1)}((1,1))+\frac{2}{5}P_{(2,0)}((1,1)).
\]
Thus
\[
L_A(p_{(1,1)})=1-\frac{2}{5}-\frac{1}{5}-\frac{2}{5}=0.
\]
Hence \(L_A(p_j)=0\) for every \(j\in S^*(C)\). Therefore the spectral condition in Theorem~\ref{clique-colique bound} is satisfied, in agreement with \(|C||A|=|X|\).\\
}

The preceding example arises from a perfect code together with a metric ball as its corresponding anticode. To illustrate that Theorem~\ref{clique-colique bound} applies more generally to clique--coclique equality beyond this ball setting, we now consider a diameter-perfect code whose corresponding optimal anticode is not a ball.

{\ex \label{example5}
Consider the threefold product scheme
\[
\mathfrak{X}=K_2\otimes K_2\otimes K_2
\]
on $X=\mathbb{F}_2^3$, with relation index set $D=\{0,1\}^3$. Let \(C=\{(0,0,0),(1,1,1)\}\), then $C$ is a $[3,1,3]_2$ MDS code and hence is diameter-perfect by \cite[Lemma~2]{Lloy}.
The corresponding optimal anticode may be taken as
\[
A=\{0\}\times\mathbb{F}_2^2=\{(0,0,0),(0,0,1),(0,1,0),(0,1,1)\}.
\]
Indeed,
\[
d_{\min}(C)=3,~ \operatorname{diam}(A)=2,~ |C||A|=2\cdot4=8=|X|.
\]
Moreover, $A$ is not a Hamming ball, since every point of $A$ has another point of $A$ at distance $2$, whereas the center of a radius-$1$ ball is at distance at most $1$ from every point of the ball.
	
Set \(M=\{(0,0,0),(1,1,1)\}\), then $C$ is an $M$-clique and $A$ is an $\overline{M}$-clique, where
\(\overline{M}=D\setminus\{(1,1,1)\}\). The primitive eigenspaces of $\mathfrak{X}$ are indexed by
$u\in\mathbb{F}_2^3$ through the characters \(\psi_u(x)=(-1)^{u\cdot x}\).
Since $C=\langle(1,1,1)\rangle$, its Fourier support is
\[
C^\perp=\{(0,0,0),(1,1,0),(1,0,1),(0,1,1)\},
\]
and hence
\[
S^*(C)=\{(1,1,0),(1,0,1),(0,1,1)\}.
\]
For $u=(u_1,u_2,u_3)$, the corresponding spectral point is
\[
p_u=\big((-1)^{u_1},(-1)^{u_2},(-1)^{u_3}\big).
\]
For the anticode $A$, the only nonzero entries of its inner distribution are
\[
a_{(0,0,0)}=a_{(0,0,1)}=a_{(0,1,0)}=a_{(0,1,1)}=1.
\]
Since all valencies are equal to $1$ and \(v_\alpha(x_1,x_2,x_3)=x_1^{\alpha_1}x_2^{\alpha_2}x_3^{\alpha_3}\), we obtain
\[
L_A(x_1,x_2,x_3)=1+x_2+x_3+x_2x_3=(1+x_2)(1+x_3).
\]
Now
\[
p_{(1,1,0)}=(-1,-1,1),~ p_{(1,0,1)}=(-1,1,-1),~ p_{(0,1,1)}=(1,-1,-1),
\]
and therefore
\[
L_A(p_{(1,1,0)})=L_A(p_{(1,0,1)})=L_A(p_{(0,1,1)})=0.
\]
Thus the spectral condition in Theorem~\ref{clique-colique bound} agrees with the diameter-perfect equality $|C||A|=|X|$.\\
}

Example~\ref{example5} extends naturally to arbitrary MDS codes in product schemes, as follows.

{\coro
Let $\mathfrak{X}=K_q^{\otimes n}$ on $X=\mathbb{F}_q^n$, and let $C\subseteq X$ be an MDS code with minimum distance $d$. For $T\subseteq\{1,\ldots,n\}$ with $|T|=d-1$, set \(A_T=\{x\in\mathbb{F}_q^n:x_i=0\text{ for }i\notin T\}\), then
\[
L_{A_T}(x_1,\ldots,x_n)=\prod_{i\in T}(1+x_i),
\]
and
\[
L_{A_T}(p_j)=\prod_{i\in T}(1+p_{j,i})=0,~ j\in S^*(C),
\]
where $p_j=(p_{j,1},\ldots,p_{j,n})$.
}
\begin{proof}
Since $C$ is MDS, \(|C|=q^{\,n-d+1}\), while \(|A_T|=q^{d-1},~ \operatorname{diam}(A_T)=d-1\).
Hence $C$ is an $M$-clique and $A_T$ is an $\overline{M}$-clique. Moreover, \(|C||A_T|=q^n=|X|\).
For $\alpha\in D=\{0,1\}^n$,
\[
v_\alpha(x_1,\ldots,x_n)=\prod_{i=1}^n x_i^{\alpha_i},~k_\alpha=(q-1)^{|\alpha|_1},
\]
and the inner distribution of $A_T$ satisfies
\[
a_\alpha=
\begin{cases}
k_\alpha, & \operatorname{supp}(\alpha)\subseteq T,\\
	   0, & \text{otherwise}.
\end{cases}
\]
Therefore,
\[
L_{A_T}(x_1,\ldots,x_n)=\sum_{S\subseteq T}\prod_{i\in S}x_i=\prod_{i\in T}(1+x_i).
\]
The second assertion follows from Theorem~\ref{clique-colique bound}.
\end{proof}

\section{The width}
We next consider a width bound in the multivariate $P$-polynomial setting. In the classical metric case, Brouwer, Godsil, Koolen, and Martin \cite{width} proved that a nonempty subset $C$ with width $w$ and dual degree $s^*$ satisfies \(w+s^*\geq d\), where $d$ is the diameter of the underlying distance-regular graph, and that equality implies that $C$ is completely regular. We now extend this result to the present multivariate setting.

For a nonempty subset $C\subseteq X$, define its \textbf{width} by
$$w(C)=\max_{x,y\in C}\widetilde d(x,y).$$
Equivalently, in terms of the inner distribution $a(C)=(a_\alpha)_{\alpha\in\mathcal D}$, $$w(C)=\max\{|\alpha|_1:a_\alpha\neq 0\},$$
since $a_\alpha=\frac{1}{|C|}\chi_C^TA_\alpha\chi_C=\frac{1}{|C|}\big|\{(x,y)\in C\times C:d_m(x,y)=\alpha\}\big|$.
Define the \textbf{diameter} by
\[
D=\max_{x,y\in X}\widetilde d(x,y)=\max_{\alpha\in\mathcal D}|\alpha|_1.
\]
For $x\in X$, let \(\widetilde d(x,C)=\min_{c\in C}\widetilde d(x,c)\), and define the \textbf{covering radius} of $C$ by
\[
\rho(C)=\max_{x\in X}\widetilde d(x,C).
\]
For a nonempty subset $C\subseteq X$, let
\[
C_i=\{x\in X:\operatorname{dist}_G(x,C)=i\},~0\leq i\leq \rho(C).
\]
Recall that $C$ is called a \textbf{completely regular code} in $G$ if its distance partition \(X=C_0\sqcup C_1\sqcup\cdots\sqcup C_{\rho(C)}\) is equitable, that is, for every $i,j$, the number of neighbours in $C_j$ of a vertex $x\in C_i$ is independent of the choice of $x\in C_i$ \cite{completely regular}.

{\lem \label{L^1 compatible}
Let $\mathfrak X=(X,\{R_\alpha\}_{\alpha\in\mathcal D})$ be an $m$-variate $P$-polynomial association scheme, and assume that the monomial order is $L^1$-compatible, then for all $x,y\in X$, \(\widetilde d(x,y)=\operatorname{dist}_G(x,y)\).
}
\begin{proof}
By the definition of $d_m(x,y)$, there exists a walk $\xi$ from $x$ to $y$ such that
\(\ell_m(\xi)=d_m(x,y)\). Hence the ordinary length of $\xi$ is
\(|\ell_m(\xi)|_1=|d_m(x,y)|_1=\widetilde d(x,y)\), and therefore \(\operatorname{dist}_G(x,y)\leq \widetilde d(x,y)\).

Conversely, let $u,v$ be adjacent vertices joined by an edge of colour $i$, then
\(d_m(u,v)\leq \epsilon_i\). Since $d_m(u,v)\in\mathcal D$ and $\epsilon_i=\epsilon_i+\boldsymbol{0}\in\mathcal D+\mathcal D$, the $L^1$-compatibility of the monomial order gives \(|d_m(u,v)|_1\leq |\epsilon_i|_1=1\).
As $u\neq v$, we have $d_m(u,v)\neq \boldsymbol{0}$, and hence \(\widetilde d(u,v)=1\).
By Proposition~\ref{weakly metric}, $\widetilde d$ satisfies the triangle inequality. Therefore, if \(x=u_0,u_1,\ldots,u_L=y\) is a shortest path in $G$, where $L=\operatorname{dist}_G(x,y)$, then
\[
\widetilde d(x,y)
\leq
\sum_{j=1}^{L}\widetilde d(u_{j-1},u_j)
=L
=\operatorname{dist}_G(x,y).
\]
Combining the two inequalities yields \(\widetilde d(x,y)=\operatorname{dist}_G(x,y)\).
\end{proof}

{\thm \label{width}
Under the assumptions of Lemma~\ref{L^1 compatible}, every nonempty subset $C\subseteq X$ satisfies \(w(C)+r(C)\geq D\). If equality holds, then $C$ is completely regular.
}
\begin{proof}
Write \(w=w(C),~ \rho=\rho(C),~ \text{and}~r=r(C)\).
Since $\{A_\alpha\}_{\alpha\in\mathcal D}$ and $\{E_j\}_{i\in J}$ are two bases of the Bose-Mesner algebra, \(\operatorname{Col}(B)=\operatorname{Span}\{E_j\chi_C:j\in J\}\).
The nonzero vectors $E_j\chi_C$ are mutually orthogonal, and $E_0\chi_C=(|C|/|X|)\mathbf 1_X\neq0$, hence \(\operatorname{rank}B=r+1\).
For \(0\leq t\leq D\), define
\[
\widehat B_t=\sum_{\substack{\alpha\in\mathcal D\\|\alpha|_1=t}}B_\alpha,~ \widehat B=(\widehat B_0,\ldots,\widehat B_D),
\]
then \((\widehat B_t)_x=|\{c\in C:\widetilde d(x,c)=t\}|\), and since $\widehat B$ is obtained from $B$ by aggregating columns, there \(\operatorname{rank}\widehat B\leq r+1\).
Choose $x\in X$ and $c\in C$ such that $\widetilde d(x,C)=\widetilde d(x,c)=\rho$. Since $\widetilde d=\operatorname{dist}_G$, let $c=u_0,u_1,\ldots,u_\rho=x$ be a shortest path, then $\widetilde d(c,u_i)=i$ and $\widetilde d(u_i,x)=\rho-i$.
Since $c\in C$, we have $\widetilde d(u_i,C)\leq i$, while
\(\rho\leq\widetilde d(x,u_i)+\widetilde d(u_i,C)=\rho-i+\widetilde d(u_i,C)\).    Thus $\widetilde d(u_i,C)=i$ for every $0\leq i\leq\rho$.
Choose $x_i\in X$ with $\widetilde d(x_i,C)=i$, then $(\widehat B_t)_{x_i}=0$ for $t<i$, while $(\widehat B_i)_{x_i}>0$. Hence the submatrix of $\widehat B$ formed by the rows $x_0,\ldots,x_\rho$ and the columns $0,\ldots,\rho$ is upper triangular with nonzero diagonal. Therefore $\operatorname{rank}\widehat B\geq\rho+1$, and then
$\rho\leq r$.
Choose $\beta\in\mathcal D$ with $|\beta|_1=D$, and fix $y\in C$. Since $R_\beta$ has positive valency, there exists $x\in X$ such that $(y,x)\in R_\beta$, so $\widetilde d(y,x)=D$. For every $c\in C$, \(D\leq\widetilde d(y,c)+\widetilde d(c,x)\leq w+\widetilde d(c,x)\). Hence $\widetilde d(x,C)\geq D-w$, and therefore \(\rho\geq D-w\).
Combining these inequalities gives \(D-w\leq\rho\leq r\), and hence $w+r\geq D$.

If $w+r=D$, then $\rho=r=D-w$. Since $\widetilde d(x,C)\geq D-w=\rho$, we have $\widetilde d(x,C)=\rho$. Choose $z\in C$ such that $\widetilde d(z,x)=\rho$, then
\(D\leq\widetilde d(y,z)+\widetilde d(z,x)\leq w+\rho=D\), so $\widetilde d(y,z)=w$.
Let $z=z_0,z_1,\ldots,z_\rho=x$ be a shortest path, then $\widetilde d(z,z_i)=i$ and $\widetilde d(z_i,x)=\rho-i$. Moreover \(\widetilde d(y,z_i)\leq w+i\), while \(w+\rho=D\leq \widetilde d(y,z_i)+\rho-i\), we have \(\widetilde d(y,z_i)=w+i\).
As $z\in C$, $\widetilde d(z_i,C)\leq i$. On the other hand, for every $c\in C$,
\(w+i\leq\widetilde d(y,c)+\widetilde d(c,z_i)\leq w+\widetilde d(c,z_i)\), so $\widetilde d(z_i,C)=i$.
Set $b_i=B(z_i)$ and $\widehat b_i=\widehat B(z_i)$. If $u\in C_i$ and $c_0\in C$ satisfies $\widetilde d(u,c_0)=i$, then for every $c\in C$, \(i\leq \widetilde d(u,c)\leq i+w\). Hence
\[
(\widehat B_t)_u=0,~\text{for }t<i\text{ or }t>i+w.\tag{5}
\]
For the special vertices $z_i$, (5) gives
\[
(\widehat B_i)_{z_i}>0,~(\widehat B_{i+w})_{z_i}>0.\tag{6}
\]
The rows $\widehat b_0,\ldots,\widehat b_\rho$ are linearly independent, since their restrictions to the columns $0,\ldots,\rho$ form an upper triangular matrix with nonzero diagonal. Therefore $b_0,\ldots,b_\rho$ are also linearly independent.
By \(\operatorname{rank}B=r+1\) and $\rho=r$, they form a basis of $\operatorname{Row}(B)$.
Let $u\in C_\ell$. Write \(B(u)=\sum_{i=0}^{\rho}\lambda_i b_i\), and then
\[
\widehat B(u)=\sum_{i=0}^{\rho}\lambda_i\widehat b_i.\tag{7}
\]
If some $\lambda_i\neq0$ with $i<\ell$, choose the smallest such index $k$. Looking at the $k$-th coordinate of (7), (5)-(6) gives $0=\lambda_k(\widehat B_k)_{z_k}$, a contradiction. Hence $\lambda_i=0$ for $i<\ell$. Similarly, if some $\lambda_i\neq0$
with $i>\ell$, choose the largest such index $k$ and look at the $(w+k)$-th coordinate. Since $w+k\leq w+\rho=D$, (5)-(6) again give a contradiction. Thus $B(u)=\lambda_\ell b_\ell$.
Every row of $B$ has row sum $|C|$, so $\lambda_\ell=1$. Therefore $B(u)=B(z_\ell)$, and hence
\[
\widetilde d(u,C)=\widetilde d(v,C)~\Longrightarrow~B(u)=B(v).\tag{8}
\]
Let $H=(\chi_{C_0},\ldots,\chi_{C_\rho})$. By (8), every column of $B$ is constant on each $C_i$, so $\operatorname{Col}(B)\subseteq\operatorname{Col}(H)$. Since both spaces have dimension $\rho+1$, they are equal. For every $\alpha\in\mathcal D$, the Bose--Mesner algebra is closed under multiplication,
so \(A_\alpha\operatorname{Col}(B)\subseteq\operatorname{Col}(B)\).
Hence $A_\alpha\operatorname{Col}(H)\subseteq\operatorname{Col}(H)$, which means that the distance partition is equitable with respect to $A_\alpha$. Since this holds for every $\alpha\in\mathcal D$, and \(A(G)=\sum_{\substack{\alpha\in\mathcal D\\|\alpha|_1=1}}A_\alpha\), the distance partition is also equitable with respect to $A(G)$.
Thus $C$ is completely regular in $G$.
\end{proof}	

We illustrate Theorem~\ref{width} with three examples. The first is a small example showing explicitly how the parameters in the theorem are computed, the second gives an infinite family in product Hamming schemes for which equality holds, and the third is a completely regular Lee-metric code in a four-dimensional lattice graph taken from \cite{CRC}.

{\ex
Consider the bivariate product scheme
\[
\mathfrak X=H(1,2)\otimes H(1,2)
\]
on $X=\{0,1\}^2$, with relation indices $\mathcal D=\{(0,0),(1,0),(0,1),(1,1)\}$. Here
$\widetilde d$ is the ordinary Hamming distance, and hence $D=2$.
Let
\[
C=\{0\}\times\{0,1\}=\{(0,0),(0,1)\},
\]
 then $w(C)=1$ and $\rho(C)=1$. With the vertex order $(0,0),(0,1),(1,0),(1,1)$,
\[
\chi_C=\chi_{\{0\}}\otimes\mathbf 1=\begin{pmatrix}1\\0\end{pmatrix}\otimes\begin{pmatrix}1\\1\end{pmatrix}=\begin{pmatrix}1\\1\\0\\0\end{pmatrix}.
\]
If $E_0,E_1$ are the primitive idempotents of $H(1,2)$, then the primitive idempotents of $\mathfrak X$ are \(E_{i,j}=E_i\otimes E_j,~ i,j\in\{0,1\}\). Since \(\chi_C=\chi_{\{0\}}\otimes\mathbf 1\), we have
\[
E_{i,j}\chi_C=(E_i\otimes E_j)(\chi_{\{0\}}\otimes\mathbf 1)
=(E_i\chi_{\{0\}})\otimes(E_j\mathbf 1).
\]
Now $E_0\mathbf 1=\mathbf 1$ and $E_1\mathbf 1=0$, while $E_0\chi_{\{0\}}\neq0$ and $E_1\chi_{\{0\}}\neq0$. Hence
\[
E_{0,0}\chi_C\neq0,~E_{1,0}\chi_C\neq0,~\text{whereas}~E_{0,1}\chi_C=E_{1,1}\chi_C=0.
\]
After excluding the trivial index $(0,0)$, \(S^*(C)=\{(1,0)\}\), and then $r(C)=1$.
Therefore, \(w(C)+r(C)=2=D\). The distance partition is
$C_0=C$ and $C_1=\{(1,0),(1,1)\}$. Every vertex has one neighbour in each of $C_0$ and $C_1$, so this partition is equitable. Thus $C$ is completely regular.
}

{\ex
Consider the $m$-variate product Hamming scheme
\[
\mathfrak X=H(n_1,q_1)\otimes\cdots\otimes H(n_m,q_m),
\]
with vertex set $X=X_1\times\cdots\times X_m$, where $X_i=Q_i^{n_i}$ and $|Q_i|=q_i$. Its relations are indexed by $\alpha=(a_1,\ldots,a_m)$, where $a_i$ is the Hamming distance in the $i$-th block. Hence $\widetilde d(x,y)=a_1+\cdots+a_m$, and \(D=\sum_{i=1}^m n_i\).
Fix $x_0\in X_1$ and let
\[
C=\{x_0\}\times X_2\times\cdots\times X_m.
\]
Since only the last $m-1$ blocks vary inside $C$, \(w(C)=\sum_{i=2}^m n_i\).
Moreover, for $x=(x_1,\ldots,x_m)\in X$, the last $m-1$ blocks can be matched exactly by a codeword of $C$, so $\widetilde d(x,C)=d_H(x_1,x_0)$. Therefore $\rho(C)=n_1$.
The characteristic vector of $C$ is
\[
\chi_C=\chi_{\{x_0\}}\otimes\mathbf 1_{X_2}\otimes\cdots\otimes\mathbf 1_{X_m}.
\]
If $E_{j_i}^{(i)}$, $0\le j_i\le n_i$, are the primitive idempotents of $H(n_i,q_i)$, then those of $\mathfrak X$ are
\[
E_{\mathbf j}=E_{j_1}^{(1)}\otimes\cdots\otimes E_{j_m}^{(m)},~\mathbf j=(j_1,\ldots,j_m).
\]
Hence
\[
E_{\mathbf j}\chi_C=\bigl(E_{j_1}^{(1)}\chi_{\{x_0\}}\bigr)\otimes\bigl(E_{j_2}^{(2)}\mathbf 1_{X_2}\bigr)\otimes\cdots\otimes\bigl(E_{j_m}^{(m)}\mathbf 1_{X_m}\bigr).
\]
For $k\ge2$, we have $E_{0}^{(k)}\mathbf 1_{X_k}=\mathbf 1_{X_k}$ and $E_{j_k}^{(k)}\mathbf 1_{X_k}=0$ for $j_k>0$. On the other hand, $E_{j_1}^{(1)}\chi_{\{x_0\}}\neq0$ for every $0\le j_1\le n_1$, since
\(\left\|E_{j_1}^{(1)}\chi_{\{x_0\}}\right\|^2=\bigl(E_{j_1}^{(1)}\bigr)_{x_0x_0}
=\frac{m_{j_1}}{|X_1|}>0\), where $m_{j_1}=\operatorname{rank}(E_{j_1}^{(1)})$. Therefore
\[
S^*(C)=\{(j_1,0,\ldots,0):1\le j_1\le n_1\},
\]
and hence $r(C)=n_1$. Thus \(w(C)+r(C)=\sum_{i=2}^m n_i+n_1=D\), so equality holds in Theorem~\ref{width}.
	
Finally, since $\widetilde d(x,C)=d_H(x_1,x_0)$, the distance layers are
\[
C_i=\Gamma_i(x_0)\times X_2\times\cdots\times X_m,~0\le i\le n_1,
\]
where $\Gamma_i(x_0)=\{x_1\in X_1:d_H(x_1,x_0)=i\}$. For $x\in C_i$, the numbers of neighbours of $x$ in $C_{i-1}$, $C_i$, and $C_{i+1}$ are respectively
\[
i,~i(q_1-2)+\sum_{k=2}^m n_k(q_k-1),~(n_1-i)(q_1-1),
\]
and there are no neighbours in any other layer. These numbers depend only on $i$, so the distance partition is equitable. Hence $C$ is completely regular.
}

{\ex
Consider
\[
G_6^4=\operatorname{Cay}\bigl(\mathbb{Z}_6^4,\{\pm e_1,\pm e_2,\pm e_3,\pm e_4\}\bigr).
\]
Since $G_6=C_6$, we regard $G_6^4$ as the direct product of four $P$-polynomial association schemes arising from the cycle $C_6$, and the relation set is indexed by \(\mathcal D=\{0,1,2,3\}^4\). We equip this $4$-variate scheme with the graded lexicographic order. Since this order first compares total degree, it is $L^1$-compatible. Hence by Lemma~\ref{L^1 compatible},
\[
\widetilde d(x,y)=d_G(x,y)=\sum_{i=1}^4 d_{C_6}(x_i,y_i).
\]
Let \(R_0=\{0,1\}\subseteq\mathbb{Z}_6\). By the addition construction in \cite{CRC},
\(C=R_0\times\mathbb{Z}_6^3\) is a completely regular code in $G_6^4$. Since $\operatorname{diam}(C_6)=3$, the diameter is \(D=12\). Moreover, \(w(C)=1+3+3+3=10\).
It remains to determine the external spectral degree. Let
\[
\omega=e^{2\pi i/6},~\psi_k(x)=\omega^{kx},~0\leq k\leq5,
\]
be the characters of $\mathbb{Z}_6$. The eigenvalue of the adjacency matrix of $C_6$ corresponding to $\psi_k$ is \(\omega^k+\omega^{-k}=2\cos\frac{2\pi k}{6}\), thus the eigenvalues for $k=0,1,\ldots,5$ are \(2,1,-1,-2,-1,1\). Hence
\[
V_0=\langle\psi_0\rangle,~V_1=\langle\psi_1,\psi_5\rangle,~V_2=\langle\psi_2,\psi_4\rangle,~V_3=\langle\psi_3\rangle.
\]
For $R_0=\{0,1\}$, its Fourier coefficient at $\psi_k$ is
\[
\widehat{\chi_{R_0}}(k)=\frac{1}{6}\sum_{x\in\mathbb{Z}_6}\chi_{R_0}(x)\overline{\psi_k(x)}=\frac{1}{6}\left(\overline{\psi_k(0)}+\overline{\psi_k(1)}\right)=\frac{1}{6}(1+\omega^{-k}),
\]
which vanishes only for $k=3$. We obtain
\[
E_1\chi_{R_0}\neq0,~E_2\chi_{R_0}\neq0,~E_3\chi_{R_0}=0.
\]
Since \(\chi_C=\chi_{R_0}\otimes\mathbf{1}^{\otimes3}\), and $\mathbf{1}$ belongs to the trivial eigenspace $V_0$ of each $C_6$ factor, the last three coordinates of every index in $S^*(C)$ must be zero. Hence
\[
S^*(C)=\{(1,0,0,0),(2,0,0,0)\},~r(C)=2.
\]
Therefore
\[
w(C)=10=12-2=D-r(C),
\]
so equality holds in Theorem~\ref{width}, consistently with the complete regularity of $C$ established in \cite{CRC}.
}

\section{Conclusion}
In this paper, we have extended several classical results of Delsarte \cite{D2} in $P$-polynomial association schemes to multivariate $P$-polynomial association schemes \cite{B+}. We have used the concept of $m$-distance regular graph of
\cite{mreg}. The older concept of weakly metric schemes has been instrumental \cite{S1}. We have obtained a multivariate Lloyd theorem for perfect codes, and a linear recursion on the columns of  the outer distribution matrix. The latter result yielded a bound on the covering radius of the code. We have studied  clique-coclique equality with applications to diameter perfect codes, and derived a Lloyd theorem for them. We have derived  bounds related to the width \cite{width}. In the equality case of the width bound, the code is shown to be completely regular. Several examples of codes in weakly metric association schemes illustrate these results.

An important open problem is to use the two Lloyd theorems above to derive non-existence results for perfect codes and diameter perfect codes.

\section{Declaration of AI Use}
During the preparation and revision of this article, the authors used ChatGPT (OpenAI) to assist with language refinement, the organization and development of ideas, and improvements in clarity. All AI-assisted content was reviewed, verified, and revised by the authors, who take full responsibility for the results, the arguments, and the final form of the article.

\end{document}